\documentclass[8pt,a4]{article}
\usepackage{graphicx}
\usepackage{amssymb,amscd,amsmath}
\usepackage{wrapfig}
\usepackage{feynmf}
\usepackage{appendix}
\usepackage{slashed}
\usepackage{indentfirst}
\usepackage{wrapfig}
\usepackage{appendix}
\usepackage[usenames]{color}

\makeatletter \@addtoreset{equation}{section} \makeatother

\newcommand{\absvec}[1]{\left| \mathbf{ #1 } \right|}

\begin{document}

\renewcommand{\abstractname}{Abstract}
\renewcommand{\refname}{References}

\title{\bf ON COLLECTIVE PROPERTIES OF TURBULENT QED PLASMA}

\author{Martin Kirakosyan${}^{(a)}$, Andrei Leonidov${}^{(a)}$\footnote{Also at the Institute of Experimental and Theoretical Physics and Moscow Institute
of Physics and Technology, Moscow, Russia}, Berndt M\"{u}ller${}^{(b)}$
\\
\\
\small{\em (a) P.N. Lebedev Physical Institute, Leninsky pr. 53, 119991 Moscow, Russia} \\
\small{\em (b) Department of Physics, Duke University, Durham, NC 27708, USA }
 }

\date{}
\maketitle

\begin{abstract}
Polarization properties of turbulent stochastically inhomogeneous ultrarelativistic QED plasma are studied. It is shown that the sign of nonlinear
turbulent Landau damping corresponds to an instability of the spacelike modes and, for sufficiently large turbulent fields, to an actual
instability of a system. Modification of plasmon dispersion relations due to turbulent effects are studied.
\end{abstract}

\bigskip

\bigskip
\newpage
\section{Introduction}

Working out a quantitative description of the properties of dense strongly interacting matter produced in ultrarelativistic heavy ion collisions
presents one of the most fascinating problems in high energy physics. The simplest (albeit not unique) way of putting the experimental data from
RHIC \cite{hydroRHIC} and LHC \cite{hydroLHC} into a coherent framework is to describe the essential physics of these collisions as a
hydrodynamical expansion of primordial quark-gluon matter that, after a short transient period, reaches sufficient level of local equilibration
allowing the usage of hydrodynamics. The features of the experimentally observed energy flow, in particular the presence of a strong elliptic
flow, suggest early equilibration of the initially produced matter and small shear viscosity of the expanding fluid, see e.g. the discussion in
\cite{H05} and \cite{MSW12} devoted to RHIC and LHC results respectively .

A coherent microscopic description of multiparticle production in high energy heavy ion collisions should embrace all its stages from initial
inelasticity to free flow of final hadrons. We are still very far from developing it and focus instead on working out plausible models aiming at
providing a reasonably simple description of the particular stages of these collisions. One of the most interesting stylized features arising in a
number of models aiming to describe the evolution of the primordial dense non-Abelian matter in the weak coupling regime is the presence of
instabilities eventually resulting in a turbulent-like state of this matter.

At the most fundamental level a description of early stages of high energy nuclear collisions in the weak coupling regime is based on the idea
that large gluon density and, correspondingly, large occupation numbers of low energy gluon modes make it natural to use tree-level Yang-Mills
equations with sources in the strong field regime as a major building block for the theoretical description of ultrarelativistic nuclear
collisions. It was shown that the strongly nonisotropic tree-level gluon field configuration arising immediately after collision, the glasma
\cite{KMW95,LM06}, that initially contains purely longitudinal chromoelectric and chromomagnetic fields, is unstable with respect to
boost-noninvariant quantum fluctuations \cite{RV06}. At later stages of its evolution these instabilities were shown to drive a system towards a
state characterized by the turbulent Kolmogorov momentum spectrum of its modes \cite{FG12}. The same Kolmogorov spectrum was earlier discovered in
a simplified scalar model of multiparticle production in heavy ion collisions \cite{DEGV11,EG11}.  A possible relation between these instabilities
and low effective viscosity in expanding geometry was recently discussed in \cite{DEGV12}.

The origin of the initial glasma instabilities and the physical picture underlying the turbulent-like glasma at later stages of its evolution,
however, do still remain unclear. The usual references are here to the Weibel-type instabilities of soft field modes present both in QED and QCD
plasma and having their origin in the momentum anisotropy of hard sources  \cite{W59,M88,PS88,M93,M97,ALM03,ALMY05} and the resulting turbulent
Kolmogorov cascade \cite{AM06a,AM06b}, see also the review \cite{R09} and the recent related development in \cite{KM11,KM12,IRS11,CR11}.

Of major importance to the physics of turbulent quantum field theory that provide another important benchmark for the physics of heavy ion
collisions are also the fixed-box studies in the framework of classical statistical lattice gauge theory \cite{BSS07,BGSS09,BSS09} and a study of
the turbulent cascade in the isotropic QCD matter in \cite{MSW07}. Let us also note that there is no doubt that the genuinely stochastic nature of
the classical Yang-Mills equation \cite{KMOSTY10} should by itself play an important role in the physics of turbulent non-Abelian matter. The
precise relation is however still to be studied.

The importance of turbulent effects makes it natural to study their effects on physically important quantities like shear viscosity. The
corresponding calculation was made in \cite{ABM06,ABM07,ABM11} in a setting generalizing the one used in the earlier studies of turbulent QED
plasma \cite{Tsit,I91}, in which turbulent plasma is described as a system of hard thermal modes and the stochastic turbulent fields characterized
by some spatial and temporal correlation lengths. It was shown that plasma turbulence can serve as a natural source of the above-mentioned
anomalous smallness of viscosity of strongly interacting matter created in high energy heavy ion collisions.

The physics of turbulence, both in liquids \cite{OYS07,ZSIG07,ZSIG08,ZS10,ZSI10,ZS12} and plasma \cite{K02}, is essentially that of space-time
structures that appear at the event-by-event level and, after averaging, give rise to Kolmogorov scaling of the structure functions. The
event-by-event stochastic inhomogeneity of turbulent plasma can therefore play an important role in forming its physical properties. In the
present paper we discuss the turbulent contributions to the most fundamental physical characteristics of plasma, the properties of its collective
modes, plasmons. For simplicity we shall restrict ourselves to considering an Abelian case; the corresponding non-Abelian generalization will
appear in a separate publication \cite{KLM13}. The effects in question can broadly be described as nonlinear Landau damping \cite{I91}. One of the
most interesting effects we see is a nonlinear Landau instability for transverse plasmons at large turbulent fields, i.e. a phenomenon equivalent
to nonlinear Landau damping, but with an opposite sign of the corresponding imaginary part of the response tensor. The origin of the phenomena
considered in the paper is in the stochastic inhomogeneity of the turbulent electromagnetic fields in QED plasma; in this respect they are similar
to the phenomenon of the stochastic transition radiation \cite{T72,KL08a,KL08b}. In particular, similarly to the stochastic transition radiation,
the turbulent contributions to plasmon properteis discussed in this paper vanish in the limit of vanishing correlation length of the stochastic
turbulent fields.

\section{Turbulent polarization and plasmons}

\subsection{Turbulent QED plasma}

The present study is devoted to the properties of ultrarelativistic electron-positron plasma (for brevity, QED plasma) defined as a system of
massless charged particles, electrons and positrons, described by the distribution functions $f(p,x,q)$, where $q=1$ for electrons and $q=-1$ for
positrons, and regular and turbulent electromagnetic fields $F^R_{\mu \nu}$ and $F^T_{\mu \nu}$ respectively.

In what follows the turbulent plasma is described as a perturbation of the original equilibrium plasma characterized by the corresponding
equilibrium distribution functions $f^{\rm eq}(p,q)$\footnote{In the explicit calculations in this paper we will use for $f^{\rm eq}(p,q)$ a Fermi
thermal distribution} by weak stochastic turbulent fields $F^T_{\mu \nu}$. It is assumed that $F^T_{\mu \nu}$ belong to a Gaussian stochastic
ensemble:
\begin{equation}
\langle F_{\mu \nu}^{T}\rangle=0, \;\;\;\; \langle F^{T \mu \nu}(x)F^{T \mu^{\prime} \nu^{\prime}}(y)\rangle=K^{\mu \nu \mu^{\prime}
\nu^{\prime}}(x,y), \label{turbens}
\end{equation}
where $K^{\mu \nu \mu^{\prime} \nu^{\prime}}(x,y)$ is a basic two-point correlator characterizing the stochastic properties of the Gaussian
ensemble of turbulent fields. In the present study we shall restrict our analysis to the simplest case of turbulent plasma that is on average
stationary and homogeneous so that the two-point correlator in (\ref{turbens}) depends only on temporal $t=\vert x^0-y^0 \vert$ and spatial
$r=\sqrt{({\mathbf x}-{\mathbf y})^2}$ differences and employ the following Ansatz for the correlator of the turbulent fields \cite{ABM07}:
\begin{equation}\label{turbcor}
K^{\mu \nu \mu^{\prime} \nu^{\prime}}(x)=K_{0}^{\mu \nu \mu^{\prime} \nu^{\prime}}\exp\left[-\dfrac{t^{2}}{2\tau^{2}}-\dfrac{r^{2}}{2
a^{2}}\right].
\end{equation}

In describing the properties of the turbulent QED plasma we will neglect collisions between plasma particles but take into account their
interaction with turbulent and regular electromagnetic fields.  The role of the regular field in considering polarization properties of the
turbulent QED plasma is in ensuring the presence of a regular external perturbation $F^R_{\mu \nu}$ defining the corresponding linear response.
Therefore, within the above-described approximation, the system of equations describing the turbulent QED plasma reads
\begin{equation}\label{kinetic+maxw}
\begin{cases}
\begin{split}
& p^{\mu}\left[ \partial_{\mu}-
e q \left( F^R_{\mu \nu} + F^T_{\mu \nu} \right) \dfrac{\partial}{\partial p_{\nu}}\right]f(p,x,q)=0\\
&\partial^{\mu}\left( F^R_{\mu \nu} + F^T_{\mu \nu} \right)=j_{\nu}\\
&j_{\nu}(x)=e \sum_{q,s}\int d P p_{\nu} q f(p,x,q) \,
\end{split}
\end{cases}
\end{equation}
where $d P = d^4p \delta(p^{2})\theta(p^{0})$ is a phase space integration for ultrarelativistic particles, the summation is over spin and charge
and, following the standard assumption \cite{Tsit}, we neglect the equilibrium contribution to the regular electromagnetic field. The unperturbed
equilibrium state is electrically neutral: electrons and positrons are characterized by the same distribution functions and the current
$j_\nu^{\rm eq}$ is absent:
\begin{equation}
j_\nu^{\rm eq} (x) =e \sum_{q,s} \int d P p_{\nu} q f^{\rm eq}(p,x,q) \equiv e \int d P p_{\nu} (f^{\rm eq}(p,q=1)-f^{\rm eq}(p,q=-1)) =0
\end{equation}

In the linear response approximation the polarization properties of the turbulent QED plasma are fully characterized by the polarization tensor
$\Pi^{\mu \nu} (k) \equiv \Pi^{\mu \nu} (\omega, \absvec{k})$ defined by the functional derivative of the average induced current $\langle
j^{\mu}(k \vert F^R,F^T) \rangle_{F^T}$ over the regular perturbation $A^R_{\nu}$
\begin{equation}
\Pi^{\mu \nu} (k) = \frac{ \delta \langle   j^{\mu}(k \vert F^R,F^T) \rangle_{F^T}}{\delta A^R_{\nu}} \label{PolarTensDef},
\end{equation}
where the induced current is determined by the corresponding field-dependent contributions to the distribution functions $\delta f (p,k,q \vert
F^R,F^T)$
\begin{equation}
\langle   j^{\mu}( k \; \vert F^R,F^T) \rangle_{F^T} = e \sum_{q,s} \int d P p_{\nu} q \langle \delta f (p,k,q \vert F^R,F^T) \rangle_{F^T}
\label{indturbcur}
\end{equation}
which are to be found by solving the kinetic equations (\ref{kinetic+maxw}). The first equation of (\ref{kinetic+maxw}) is conveniently rewritten
in momentum representation as
\begin{equation}
 f(p,k,q\; \vert F^R,F^T)=f^{\rm eq}(p,q)\delta^{4}(k) - \imath \dfrac{e \,q \, p^{\mu}}{(p k)+ \imath \epsilon}\int d^{4}\,k_{1}\,
 F_{\mu \nu}(k-k_1) \dfrac{\partial f(p,k_1,q \; \vert F^R,F^T)}{\partial p_{\nu}}, \label{kineqmr}
\end{equation}
where  the term $\imath \epsilon $ in the denominator of propagator corresponds to choosing the Landau retarded boundary conditions.

To elucidate the structure of the forthcoming calculation let us first rewrite Eq.~(\ref{kineqmr}) in the following compact form:
\begin{equation} 
f=f^{eq} + G p^{\mu}F_{\mu\nu}\partial_{p}^{\mu}f \; ,
\end{equation}\label{kineqcondensed}
where $G$ is a free kinetic propagator
\begin{equation}
\imath G \equiv \dfrac{e q}{(pk)+\imath \epsilon}
\end{equation}

The linear response approximation we employ means that we are interested in the correction to the distribution functions which is of the first
order in the regular perturbation $F^R_{\mu \nu}$ and the lowest order nontrivial turbulent contribution comes from the terms which are of the
second order in the turbulent field $F^T_{\mu \nu}$, see Eq.~(\ref{turbens}). Therefore the turbulent contribution for induced current comes from
the cubic terms that are of the first order in $F^R_{\mu \nu}$ and of the second order in $F^T_{\mu \nu}$. To extract the corresponding terms it is convenient to introduce a formal expansion in parameters $\rho$ and $\tau$ counting the powers of the regular and turbulent electromagnetic fields respectively:
\begin{eqnarray}
f & = & \sum_{m=0}\sum_{n=0}\rho^{m}\tau^{n}\delta f_{mn} \label{expdisfun} \\
F^{\mu \nu}& = & \sum_{m=0}\sum_{n=0}\rho^{m}\tau^{n}F_{mn}^{\mu\nu}, \label{expfield}
\end{eqnarray}
where $\delta f_{00} = f^{\rm eq}$, $F_{00}^{\mu\nu}=0$, $F_{10}^{\mu\nu}=F^R_{\mu \nu}$ and $F_{01}^{\mu\nu}=F^T_{\mu \nu}$.
The explicit expressions for $\delta f_{mn}$ follow from substituting the expansions (\ref{expdisfun},\ref{expfield}) into the kinetic equation (\ref{kineqcondensed}) and comparing, order by order in $\rho^{m}\tau^{n}$, contributions to the left- and right- hand side, while those for $F_{mn}^{\mu\nu}$ follow from substituting (\ref{expdisfun},\ref{expfield}) into the second equation of (\ref{kinetic+maxw}).

The leading non-turbulent contribution corresponds to computing polarization in the Hard Thermal Loop (HTL) approximation:
\begin{equation}
\delta f_{\rm HTL} \equiv \delta f_{10}=  Gp_{\mu}F_{10}^{\mu\nu}\partial_{\mu,p}f^{\rm eq},
\end{equation}\label{fdistHTL}
while the leading turbulent contribution is
\begin{equation}
\delta f_{12}=Gp_{\mu}(F_{01}^{\mu \nu}\partial_{\nu, p}\;\delta f_{11}+F_{11}^{\mu \nu}\partial_{\nu, p}\;\delta f_{01}+F_{12}^{\mu
\nu}\partial_{\nu,p} f^{\rm eq}+F_{10}^{\mu \nu}\partial_{\nu, p}\;\delta f_{02})
\label{fdistturb0}
\end{equation}
The contributions in (\ref{fdistturb0}) proportional to $F_{11}$ and $F_{12}$ can be shown to be subleading. The last term in (\ref{fdistturb0}) proportional to $\delta f_{02}$ is omitted because of the simplifying assumption of approximate stationarity of the distribution of hard particles, see \cite{Tsit} and a discussion in the Appendix~\ref{Dupree}. Thus we are left with
\begin{equation}
\delta f_{12}=Gp_{\mu} F_{01}^{\mu \nu}\partial_{\nu, p}\;\delta f_{11} \simeq Gp_{\mu} F_{01}^{\mu \nu}\partial_{\nu, p}\;
G p_{\mu'} \left(F_{10}^{\mu' \nu'} \partial_{\nu'} \delta f_{01} + F_{01}^{\mu' \nu'} \partial_{\nu'} \delta f_{10} \right)
\end{equation}\label{fdistturb1}
Using the explicit expressions $\delta f_{01}=Gp_{\mu}F_{01}^{\mu\nu}\partial_{\nu,p}f^{\rm eq}$ and $\delta f_{10}=Gp_{\mu}F_{10}^{\mu\nu}\partial_{\mu,p}f^{\rm eq}$, we get
\begin{equation}
 \delta f_{12}\approx Gp_{\mu} \left[ F_{01}^{\mu\nu}\partial_{\nu,p}Gp_{\mu^{\prime}} F_{10}^{\mu^{\prime}\nu^{\prime}}\partial_{\nu^{\prime},p}Gp_{\rho} F_{01}^{\rho \sigma} +
 F_{01}^{\mu \nu}\partial_{\nu,p}Gp_{\mu^{\prime}} F_{01}^{\mu^{\prime}\nu^{\prime}}\partial_{\nu^{\prime},p}Gp_{\rho} F_{10}^{\rho \sigma} \right ]
 \partial_{\sigma,p} f^{\rm eq} \; ,
\label{f12}
\end{equation}
that is to be used in computing the turbulent contributions to the (averaged) induced current and the polarization operator
(\ref{PolarTensDef},\ref{indturbcur}). The final answer for the averaged variation of the distribution function is obtained by adding the 
HTL contribution $\delta f_{10}$ from (\ref{fdistHTL}) and the expression for $\delta f_{12}$ in (\ref{f12}) averaged over the ensemble of stochastic turbulent fields $\{ F_{01}
\}$:
\begin{equation}
\delta f  \simeq  \delta f_{\rm HTL} + \langle \delta f_{12} \rangle_{\rm I}  + \langle \delta f_{12} \rangle_{\rm II} \; ,
\end{equation}
where
\begin{eqnarray}
\delta f_{\rm HTL} & = & Gp_{\mu}F_{10}^{\mu\nu}\partial_{\mu,p}f^{\rm eq} \label{fhtl} \\
 \langle \delta f_{12} \rangle_{\rm I} & = & Gp_{\mu} \langle F_{01}^{\mu\nu}\partial_{\nu,p}Gp_{\mu^{\prime}} F_{10}^{\mu^{\prime}\nu^{\prime}}\partial_{\nu^{\prime},p}Gp_{\rho} F_{01}^{\rho
 \sigma}\rangle \partial_{\sigma,p} f^{\rm eq} \label{f12I} \\
\langle \delta f_{12} \rangle_{\rm II} & = & Gp_{\mu} \langle  F_{01}^{\mu \nu}\partial_{\nu,p}Gp_{\mu^{\prime}}
F_{01}^{\mu^{\prime}\nu^{\prime}}\partial_{\nu^{\prime},p}Gp_{\rho} F_{10}^{\rho \sigma}
 \rangle \partial_{\sigma,p} f^{\rm eq} \label{f12II}
\end{eqnarray}

\subsection{Turbulent polarization}\label{subsecturbpol}

Let us now turn to the calculation of the turbulent contribution to the polarization operator. From gauge invariance and isotropy it follows that
polarization tensor has only two independent components longitudinal and transverse, that we define as (here $k\equiv|\mathbf{k}|$):
\begin{equation}
\Pi_{i j}(\omega,\mathbf k \vert \; l)=\left(\delta_{i j}-\dfrac{k_{i}k_{j}}{k^{2}}\right)\Pi_{T}(\omega,\absvec{k} \vert \; l)
+\dfrac{k_{i}k_{j}}{k^{2}}\Pi_{L} (\omega,\absvec{k} \vert \; l) \label{poltendec}
\end{equation}
where
\begin{equation}
l \equiv \sqrt{2} \frac{\tau a}{\sqrt{\tau^2+a^2}}
\end{equation}
stands for the synthetic correlation scale of the turbulent fields\footnote{Explicit calculations discussed below show that the answer does indeed
depend only on $l$. }.

The polarization tensor (\ref{poltendec}) is related to the dielectric permittivity $\varepsilon_{i j}(\omega,\mathbf k \vert \; l)$ by
\begin{equation}
\varepsilon_{ij}(\omega,\mathbf k \vert \; l)=1-\dfrac{\Pi_{i j}(\omega,\mathbf k \vert \; l)}{\omega^{2}}, \label{epsilonij}
\end{equation}
so that
\begin{equation}
\varepsilon_{L(T)}(\omega,\mathbf k \vert \; l)=1-\dfrac{\Pi_{L(T)}(\omega,\mathbf k \vert \; l)}{\omega^{2}} \label{epsilonLT}
\end{equation}

Let us rewrite the polarization tensor $\Pi_{i j}(\omega,\mathbf k \vert \; l)$ as a sum of HTL and turbulent contributions
\begin{equation}
\Pi_{L(T)} (\omega,\mathbf{k} \vert \; l)  =   \Pi^{\; \rm HTL}_{L(T)} (\omega,\mathbf{k}) +  \Pi^{\; \rm turb}_{L(T)} (\omega,\mathbf{k} \vert \;
l) \label{poltenregturb}
\end{equation}
and, keeping the contributions of the first order in the turbulent energies  $\langle E_{\rm turb}^2 \rangle$ and $\langle B_{\rm turb}^2
\rangle$, study the gradient expansion of $\Pi^{\; \rm turb}_{L(T)} (\omega,\absvec{k})$ in the expansion parameter $(\absvec{k}l) < 1$:
\begin{equation}
\Pi^{\; \rm turb}_{L(T)} (\omega,\absvec{k} \vert \; l)  =  \sum_{n=1}^\infty \dfrac{(\absvec{k}l)^n}{\mathbf{k}^2}
 \left [ \phi^{\; (n)}_ {L(T)} \left(\dfrac{\omega}{\absvec{k}} \right) \langle E_{\rm turb}^2 \rangle +
 \chi^{\; (n)}_{L(T)} \left(\dfrac{\omega}{\absvec{k}}\right) \langle B_{\rm turb}^2 \rangle \right]
\label{gradexp}
\end{equation}
where the corresponding averaging in the isotropic case under consideration that leads to the structure of the answer shown in (\ref{gradexp}) is
performed by using
\begin{equation}
\left\langle B^{i}_{\rm turb}B^{j}_{\rm turb}\right\rangle=\dfrac{\delta^{i j}}{3}\left\langle B^{2}_{\rm turb}\right\rangle, \;\;\; \left\langle
E^{i}_{\rm turb}E^{j}_{\rm turb}\right\rangle=\dfrac{\delta^{i j}}{3}\left\langle E^{2}_{\rm turb}\right\rangle, \;\;\; \left\langle E^{i}_{\rm
turb}B^{j}_{\rm turb}\right\rangle=0,
\end{equation}

Let us turn to en explicit evaluation of the variations of the distribution functions in Eqs.~(\ref{fhtl})-(\ref{f12II}) and the corresponding
contributions to the polarization tensor. We shall calculate the turbulent contributions to the polarization tensor in the first two orders in the gradient expansion in $(\absvec{k}l)$.

\medskip

\noindent {\bf 1.} The HTL contribution in (\ref{fhtl}) reads
\begin{fmffile}{diag_23_2}
\begin{equation}
\delta f_{\rm HTL}=\parbox{20mm}{
\begin{fmfgraph*}(20,15)
\fmfleft{i1,i2,i3} \fmfright{o1,o2,o3} \fmf{phantom}{i1,o1} \fmf{dashes}{i2,v2} \fmf{plain,lab.dist=.05w,lab.side=right,lab=$eq$}{v2,o2}
\fmf{phantom,tension=+0.3}{i3,v3} \fmf{phantom}{v3,o3} \fmf{dbl_wiggly,lab.dist=.05w,lab.side=left,lab=$R$,tension=-0.01}{v2,v3} \fmfdot{v2}
\end{fmfgraph*}}=\dfrac{1}{\imath((p k)+\imath \epsilon)}\cdot e q p^{\mu}\dfrac{\partial}{\partial p_{\nu}}\cdot F_{\mu \nu}(k)f^{eq}(p,k,q)
\end{equation}
\end{fmffile}
where we have introduced self-explanatory diagram notations. The corresponding induced current is $\delta j^{\mu}_{\rm HTL} (k) = e \sum_q,s \int
dP\, p^{\mu} q \delta f_{\rm HTL}$ leading to
\begin{equation}
\Pi_{\rm HTL}^{\mu \nu}=e^{2} \dfrac{2}{(2 \pi)^{3}} \sum_{q,s} q^{2} \int d \Omega_{\mathbf{v}} \dfrac{- (v k) v^{\mu} g^{\nu 0}+k^{0}v^{\mu}
v^{\nu}}{(v k)+\imath \epsilon} \left( -\int p^{2} \dfrac{d f^{eq}(p)}{dp} d p \right) \label{SilinPolTens0}
\end{equation}
and, after performing angular and momentum integrations in (\ref{SilinPolTens0}), to the well-known expressions(\cite{Silin60}:
\begin{equation}
\begin{split}
&\Pi_{L }^{\mathrm HTL} (\omega,\absvec{k}) =- m^2_D x^2 \left[1-\dfrac{x}{2} \; L(x) \right]\\
&\Pi_{T}^{\mathrm HTL} (\omega,\absvec{k})= m^2_D \dfrac{x^2}{2} \left[1+\dfrac{1}{2 x} \; (1-x^2 )\; L(x) \right]
\end{split}
\label{SilinPolTens}
\end{equation}
where $m^2_D=e^2 T^2/3$ is a Debye mass, $x \equiv \omega/\absvec{k}$, and
\begin{equation}
L(x) \equiv \ln\left|\dfrac{1+x}{1-x}\right|-\imath\pi\theta(1-x)
\end{equation}
The imaginary part of $L(x)$ corresponds to Landau damping in collisionless plasma.

\medskip

\noindent {\bf 2.} An explicit expression for the first turbulence-induced contribution in (\ref{f12I}) reads\footnote{Some details on this
calculation are provided in the Appendix.}
\begin{fmffile}{diag20_1}
\begin{equation}
\begin{split}
\langle \delta f_{12} \rangle_{\rm I}=\parbox{35mm}{\begin{fmfgraph*}(35,15) \fmfleft{i1,i2,i3} \fmfright{o1,o2,o3} \fmf{phantom}{i1,o1}
\fmf{dashes}{i2,v1} \fmf{dashes}{v1,v2} \fmf{dashes}{v2,v3} \fmf{plain,lab.dist=.05w,lab.side=right,lab=$eq$}{v3,o2}
\fmf{phantom,tension=1.4}{i3,u1} \fmf{phantom,tension=1.5}{u1,u2} \fmf{phantom,tension=1.6}{u2,u3} \fmf{phantom,tension=9.}{u3,o3} \fmffreeze
\fmf{wiggly,lab.dist=.05w,lab.side=left,lab=$K$,left,tension=-0.5}{v1,v2}
\fmf{dbl_wiggly,lab.dist=.05w,lab.side=left,lab=$R$,tension=-0.01}{v3,u3} \fmfdot{v1} \fmfdot{v2} \fmfdot{v3}
\end{fmfgraph*}}=&\dfrac{e^{3}\,q^3}{\imath ((p k)+\imath \epsilon)}F^{R \;\rho \sigma}(k)\cdot\int d^{4}k_{1}\;p_{\mu}\dfrac{\partial}{\partial \, p^{\nu}}K^{\mu \nu \mu^{\prime}\nu^{\prime}}(k_{1})\\
&\times \dfrac{1}{\imath((p (k-k_{1}))+\imath \epsilon)} p_{\mu^{\prime}}\dfrac{\partial}{\partial \, p^{\nu^{\prime}}}\dfrac{1}{\imath ((p
k)+\imath \epsilon)} p_{\rho}\dfrac{\partial}{\partial \, p^{\sigma}}f^{eq}(p)
\end{split}
\label{AnsDiagr1}
\end{equation}
\end{fmffile}

Let us compute the first two terms in the gradient expansion (\ref{gradexp}) in Eq. (\ref{AnsDiagr1}). The answer for the leading order contributions $\phi^{\; (1)}_ {L(T)}$ and $\chi^{\; (1)}_ {L(T)}$ reads
\begin{eqnarray}
 \phi_{\rm I\;T}^{\; (1)} (x) & = &\frac{\imath e^{4}}{6\pi\sqrt{\pi}} \; 2 x \left[\dfrac{4+10 x^{2}-6 x^{4}}{3(1-x^{2})}+x(1-x^{2}) \; L(x) \right] \label{phi1T} \\
 \phi_{\rm I\;L}^{\; (1)} (x) & = & -\frac{\imath e^{4}}{6\pi\sqrt{\pi}}\; \dfrac{8 x^{3}}{3(1-x^{2})^{2}} \label{phi1L}
\end{eqnarray}
and
\begin{eqnarray}
 \chi_{\rm I\;T}^{\; (1)} (x) & = & \frac{\imath e^{4}}{6\pi\sqrt{\pi}}\; 4 x \left[ \dfrac{-2+6 x^{2}}{3(1-x^{2})}+x \; L(x)
 \right]\label{chi1T} \\
 \chi_{\rm I\;L}^{\; (1)} (x) & = & - \frac{\imath e^{4}}{6\pi\sqrt{\pi}} \; \dfrac{8 x^{3}}{3(1-x^{2})^{2}} \label{chi1L}.
\end{eqnarray}
For the second order contributions $\phi^{\; (2)}_ {L(T)}$ and $\chi^{\; (2)}_ {L(T)}$ one gets
\begin{eqnarray}
 \phi_{\rm I\;T}^{\; (2)} (x) & = & \frac{e^{4}}{6\pi^{2}} \; x \left[\dfrac{22}{3}x+4 x^{3}  +(1+3 x^{2}+2 x^{4} ) \; L(x)
 \right]\label{phi21T} \\
 \phi_{\rm I\;L}^{\; (2)} (x) & = &\frac{e^{4}}{6\pi^{2}}\; 2 x^{3} \left[\dfrac{2 x}{1-x^{2}}+L(x)\right] \label{phi21L}
\end{eqnarray}
and
\begin{eqnarray}
 \chi_{\rm I\;T}^{\; (2)} (x) & = & \frac{e^{4}}{6\pi^{2}} \; x \left[14 x+(1- 7 x^{2} ) \; L(x)
 \right]\label{chi21T} \\
 \chi_{\rm I\;L}^{\; (2)} (x) & = &\frac{e^{4}}{6\pi^{2}}\; 2 x \left[\dfrac{6 x-4 x^{3}}{1-x^{2}}+(1-2 x^{2}) \; L(x)\right] \label{chi21L}
\end{eqnarray}

\medskip

\noindent {\bf 3.} Let us now turn to the computation of the second turbulent contribution (\ref{f12II})

\medskip

\begin{fmffile}{diag_form_1}
\begin{equation}
\begin{split}
\langle \delta f_{12} \rangle_{\rm II} =\parbox{35mm}{
\begin{fmfgraph*}(35,15)
\fmfleft{i1,i2,i3} \fmfright{o1,o2,o3} \fmf{phantom}{i1,o1} \fmf{dashes}{i2,v1} \fmf{dashes}{v1,v2} \fmf{dashes}{v2,v3}
\fmf{plain,lab.dist=.05w,lab.side=right,lab=$eq$}{v3,o2} \fmf{phantom,tension=1.4}{i3,u1} \fmf{phantom,tension=1.5}{u1,u2}
\fmf{phantom,tension=1.6}{u2,u3} \fmf{phantom,tension=9.}{u3,o3} \fmf{phantom,tension=1.3}{i1,k1} \fmf{phantom,tension=1.}{k1,o1} \fmffreeze
\fmf{wiggly,lab.dist=.05w,lab.side=left,lab=$K$,left,tension=-0.3}{v1,v3}
\fmf{dbl_wiggly,lab.dist=.05w,lab.side=left,lab=$R$,tension=-0.01}{v2,k1} \fmfdot{v1} \fmfdot{v2} \fmfdot{v3}
\end{fmfgraph*}}=&\dfrac{e^{3}\,q^3}{\imath ((p k)+\imath \epsilon)}F^{R \; \rho \sigma}(k)\cdot\int d^{3}k_{1} K^{\mu \nu \mu^{\prime}\nu^{\prime}}(k_{1})p_{\mu}\dfrac{\partial}{\partial \, p^{\nu}}\\
&\times\dfrac{1}{\imath((p (k-k_{1}))+\imath \epsilon)} p_{\rho}\dfrac{\partial}{\partial \, p^{\sigma}}\dfrac{1}{\imath(-(p k_{1})+\imath
\epsilon)}p_{\mu^{\prime}}\dfrac{\partial}{\partial \, p^{\nu^{\prime}}}f^{eq}(p) \label{AnsDiagr2}
\end{split}
\end{equation}
\end{fmffile}
An inspection of (\ref{AnsDiagr2}) shows that all terms of the first order in the gradient expansion are absent,
\begin{equation}
\phi^{ (1\;)}_ {\rm II\;L(T)} = \chi^{ (1\;)}_ {\rm II\;L(T)} = 0,
\end{equation}
while the contribution of the second order term is purely electric,  $\chi^{ (2\;)}_ {\rm II\;L(T)}=0$.
This happens because the diagram (\ref{AnsDiagr2}) has two propagators convoluted, under integration over $k$, with the turbulent correlator. The answer for the only nontrivial contribution  $\phi^{ (2\;)}_ {\rm II\;L(T)}$ reads:
\begin{eqnarray}
 \phi_{\rm II\;T}^{\; (2)} (x) & = & \frac{e^{4}}{6\pi^{2}} \; \left[\dfrac{2}{3} x^{2}-4 x^{4}  -x(1 +x^{2}-2 x^{4}) \; L(x)
 \right]\label{phi22T} \\
 \phi_{\rm II\;L}^{\; (2)} (x) & = &\frac{e^{4}}{6 \pi^{2}}\; \left[4 x^{2}-2 x^{3}\; L(x)\right] \label{phi22L}
\end{eqnarray}

\subsection{Physical consequences.}

Let us now discuss the physical consequences resulting from the effects of turbulent polarization calculated in the previous paragraph.

\subsubsection{Turbulent instability and damping}

\begin{figure}
\begin{minipage}[b]{0.45\linewidth}
\centering
\includegraphics[height=0.25\textheight, width=0.9\textwidth]{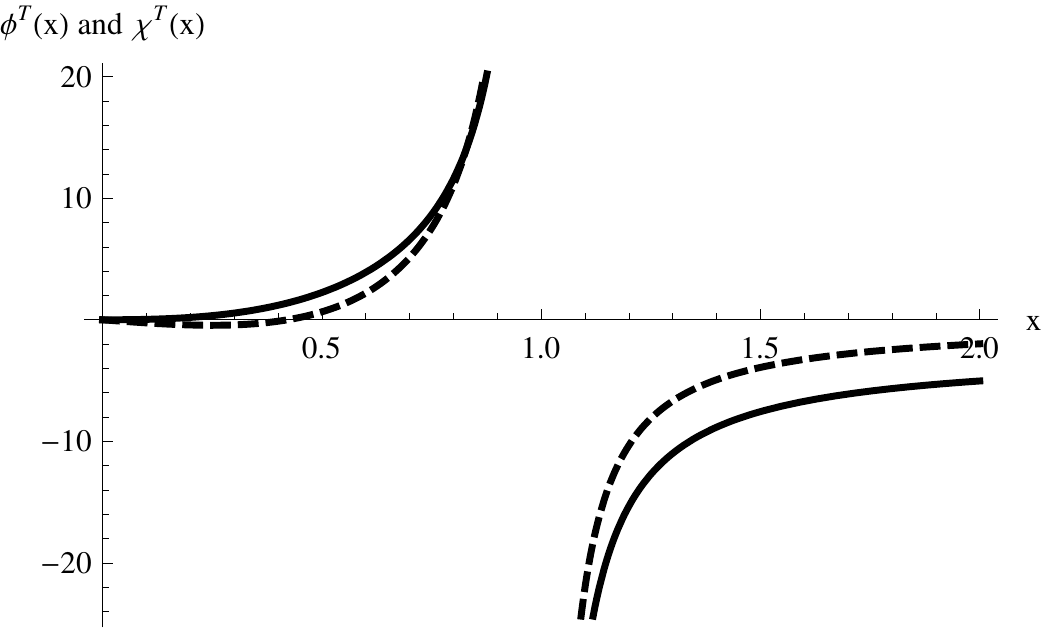}
\caption{The functions $\dfrac{6\pi \sqrt{\pi}}{e^{4}} {\mathrm{Im}}\left[\phi_T^{\; (1)} (x)\right]$ (solid line) and $\dfrac{6\pi \sqrt{\pi}}{e^{4}} {\mathrm{Im}}\left[\chi_T^{\; (1)} (x)\right]$ (dashed line).}
\label{pct}
\end{minipage}
\hfill
\begin{minipage}[b]{0.45\linewidth}
\centering
\includegraphics[height=0.25\textheight, width=0.9\textwidth]{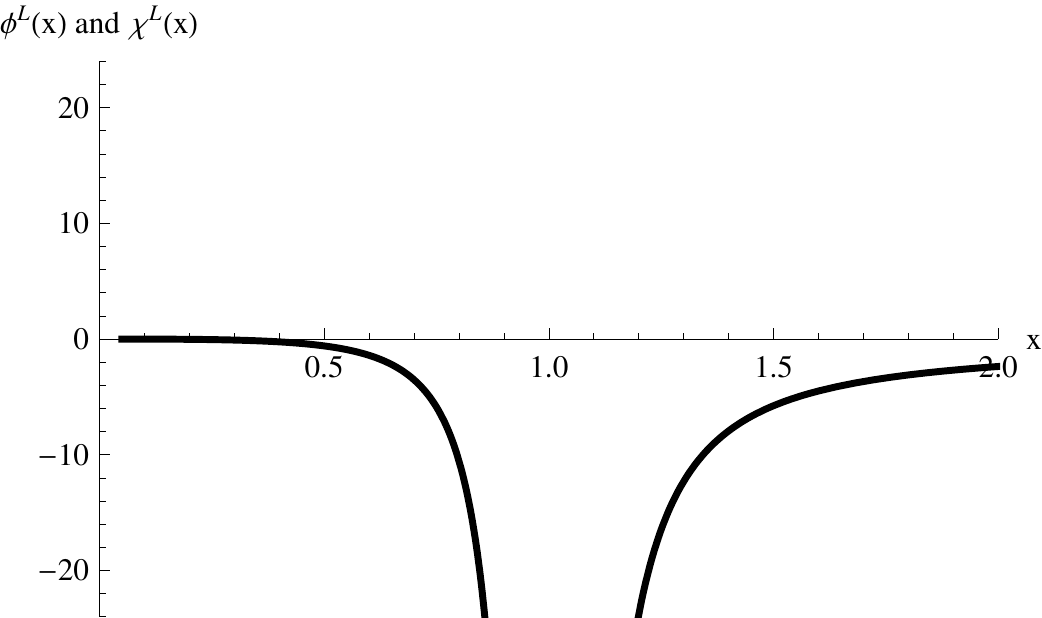}
\caption{The functions $\dfrac{6\pi \sqrt{\pi}}{e^{4}} {\mathrm{Im}}\left[\phi_L^{\; (1)} (x)\right]$ (solid line) and $\dfrac{6\pi \sqrt{\pi}}{e^{4}} {\mathrm{Im}}\left[\chi_L^{\; (1)} (x)\right]$ (dashed line).}
\label{pct1}
\end{minipage}
\end{figure}

Let us first analyze the imaginary part of the polarization operator (\ref{poltenregturb}) in the first order in the gradient expansion
(\ref{gradexp}). The analysis in the paragraph \ref{subsecturbpol} has shown that in this order the only turbulent contributions to the polarization tensor are $\phi_{\rm I\;T (L)}^{(1)}$ and $\chi_{\rm I\;T (L)}^{(1)}$ from (\ref{AnsDiagr1}), so that
\begin{eqnarray}
\mathrm{Im} \Pi_{T} (\omega,\mathbf{k} \vert \; l) &\simeq&- \pi m^2_D \dfrac{x}{4}(1-x^{2})\theta(1-x)+\dfrac{(\absvec{k}
l)}{\mathbf{k}^{2}}\left(\left\langle E^{2}\right\rangle {\rm Im} \phi_{\rm I\;T}^{(1)} (x) +
\left\langle B^{2}\right\rangle {\rm Im}\chi_{\rm I\;T}^{(1)} (x)\right) \label{impartT}\\
\mathrm{Im} \Pi_{L} (\omega,\mathbf{k} \vert \; l) &\simeq& -\pi m^2_D \dfrac{x^{3}}{2}\theta(1-x)+\dfrac{(\absvec{k} l)}{\mathbf{k}^{2}}
\left(\left\langle E^{2}\right\rangle {\rm Im} \phi_{\rm I\;L}^{(1)} (x) +\left\langle B^{2}\right\rangle {\rm Im} \chi_{\rm I\;L}^{(1)} (x)\right),
\label{impartL}
\end{eqnarray}
where the functions $\rm{Im} \phi_{\rm I\;L(T)}^{(1)}$ and  $\rm{Im} \chi_{\rm I\;L(T)}^{(1)}$ are given by Eqs.~(\ref{phi1T})-(\ref{chi1L}). The functions $\rm{Im} \phi_{\rm I\;T}^{(1)}$, $\rm{Im} \chi_{\rm I\;T}^{(1)}$ and $\rm{Im} \phi_{\rm I\;L}^{(1)}$, $\rm{Im} \chi_{\rm I\;L}^{(1)}$ are plotted in Figs. (\ref{pct}) and (\ref{pct1}) respectively.
\medskip

\noindent {\small \bf 1.} {\bf Timelike domain.} From Figs. \ref{pct} and \ref{pct1} we see that the sign of the imaginary part of the turbulent contribution to the polarization operator in the timelike domain $x>1$ is negative and corresponds to turbulent damping of timelike collective excitations. This refers to both transverse and longitudinal modes. As the HTL contribution in this domain is absent, this turbulent damping is a universal phenomenon present for all $\omega,k$ such that $\omega > k$ and all values of the parameters involved ($l$, $\langle B^2 \rangle$, $\langle E^2 \rangle$). The turbulent damping leads to an attenuation of the propagation of collective excitations at some characteristic distance.

\noindent {\small \bf 2.} {\bf Spacelike domain.} The situation in the spacelike domain $x,1$ is more diverse. In contrast with the timelike domain the gradient expansion for the imaginary part of the polarization tensor starts from the negative HTL contribution of (\ref{SilinPolTens}) corresponding to Landau damping. As seen from Figs. \ref{pct} and \ref{pct1} the imaginary parts of turbulent contributions to the longitudinal polarization tensor are negative and are thus just amplifying the Landau damping. Most interesting contributions come from turbulent contributions to transverse polarization tensor (\ref{phi1T})  and (\ref{chi1T}). From Figs. \ref{pct} and \ref{pct1} we see that the electric contribution ${\mathrm{Im}}\left[\phi_T^{\; (1)} (x)\right]$ in the spacelike domain is positive at all $x$ while the magnetic one is negative at $x<x^*$ where $x^* \sim 0.43$ and positive at $x>x^*$. This means that for sufficiently strong turbulent fields the turbulent plasma under consideration becomes unstable.

\begin{figure}
\begin{minipage}[b]{0.45\linewidth}
\centering
\includegraphics[height=0.25\textheight, width=0.9\textwidth]{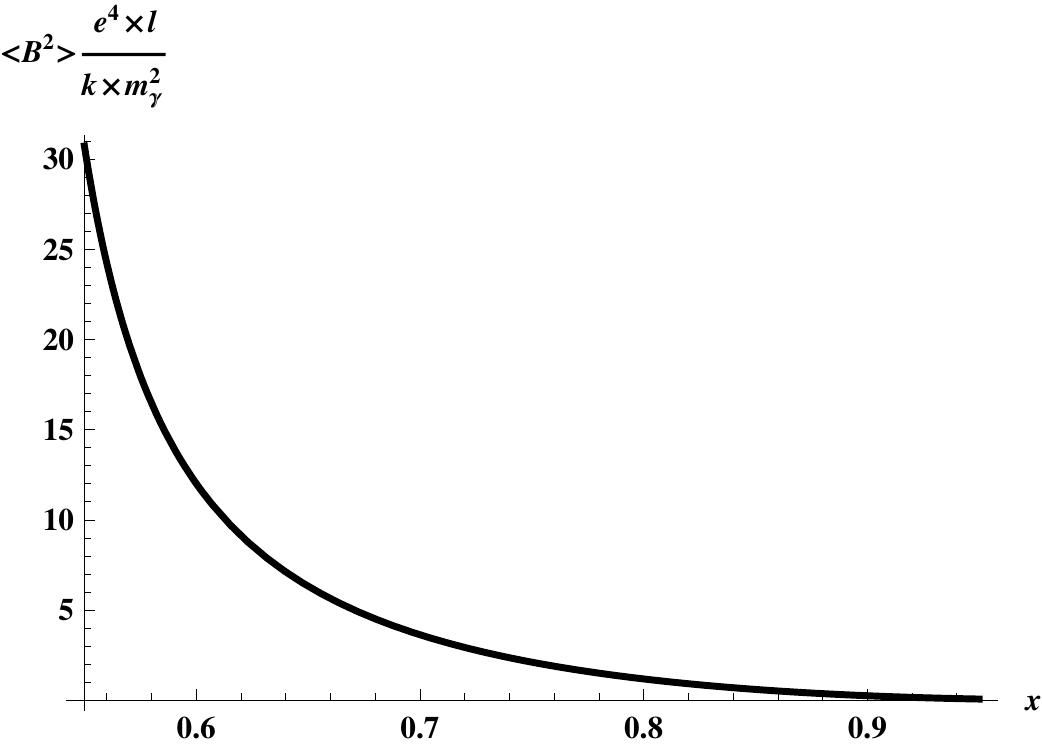}
\caption{The critical values of $e^4 (\absvec{k}l) \langle B^2 \rangle /\mathbf{k}^2 m^2_D$ for purely magnetic instability.}
\label{figminst}
\end{minipage}
\hfill
\begin{minipage}[b]{0.45\linewidth}
\centering
\includegraphics[height=0.25\textheight, width=0.9\textwidth]{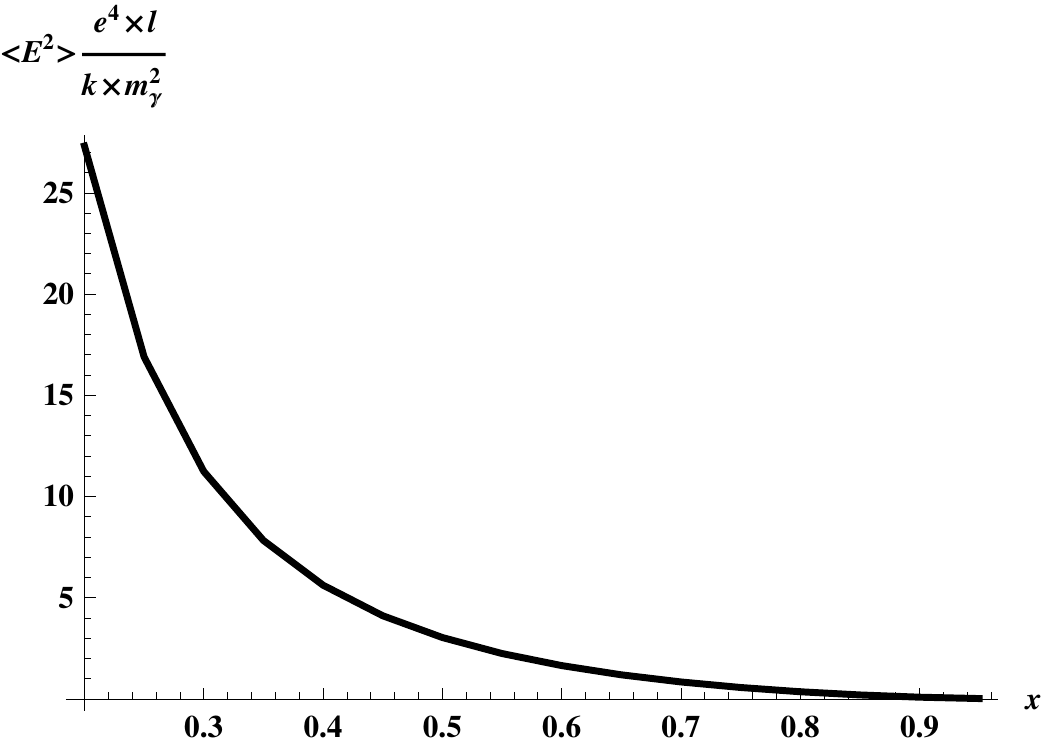}
\caption{The critical values of $e^4 (\absvec{k}l) \langle B^2 \rangle /\mathbf{k}^2 m^2_D$ for the mixed instability with $\langle B^2 \rangle = \langle E^2 \rangle$.}
\label{figeminst}
\end{minipage}
\end{figure}

Let us first study the onset of this instability in the case where turbulent fields are purely magnetic. We have
\begin{equation}
{\rm Im} \Pi_{T} (\omega,\absvec{k}) = - \pi \dfrac{e^2 T^2}{12} x(1-x^2) \left [ 1-\dfrac{4}{\pi^2 \sqrt{\pi}}
\dfrac{(\absvec{k}l)}{\mathbf{k}^2} \dfrac{e^2\langle B^2 \rangle}{T^2 } \Phi(x) \right]
\label{minst1}
\end{equation}
where
\begin{equation}
\Phi(x) = \dfrac{1}{x(1-x^2)} \left[ \dfrac{-4+12 x^2}{3(1-x^2)}+2x \; \ln\left|\dfrac{1+x}{1-x}\right| \right]
\label{minst2}
\end{equation}
From (\ref{minst1}) and (\ref{minst2}) it is clear that the condition for the appearance of the unstable regime corresponds to
\begin{equation}
\dfrac{4}{\pi^{2}\sqrt{\pi}}\dfrac{(\absvec{k}l)}{\mathbf{k}^2} \dfrac{e^2\langle B^2 \rangle}{T^2 } \Phi(x) > 1
\end{equation}
The onset of this instability is illustrated in Fig. \ref{figminst} in which we plot the critical value of the dimensionless combination $e^2 (\absvec{k}l) \langle B^2 \rangle /(\mathbf{k}^2 m^2_D)$
\medskip
For completeness let us also consider the case of a mixed instability when both electric and magnetic turbulent fields are present. For simplicity we assume that $\langle B^2 \rangle = \langle E^2 \rangle$. We get
\begin{equation}
{\rm Im} \Pi_{T} (\omega,\absvec{k}) = - \pi \dfrac{e^2 T^2}{12} x(1-x^2) \left [ 1-\dfrac{4}{\pi^2 \sqrt{\pi}}
\dfrac{(\absvec{k}l)}{\mathbf{k}^2} \dfrac{e^2\langle E^2 \rangle}{T^2 } \Theta (x) \right]
\label{minst1}
\end{equation}
where
\begin{equation}
\Theta (x) = \dfrac{1}{x(1-x^2)} \left[  \dfrac{x^2(22-6x^2)}{3(1-x^2)} + x(3-x) \; \ln\left|\dfrac{1+x}{1-x}\right| \right]
\label{minst2}
\end{equation}
so that the instability criterion reads
\begin{equation}
\dfrac{4}{\pi^{2}\sqrt{\pi}}\dfrac{(\absvec{k}l)}{\mathbf{k}^2} \dfrac{e^2\langle E^2 \rangle}{T^2 } \Theta(x) > 1
\end{equation}
and is illustrated, for the dimensionless combination $e^2 (\absvec{k}l) \langle E^2 \rangle /(\mathbf{k}^2 m^2_D)$, in the Fig. \ref{figeminst}.

As the origin of this instability is in event-by-event inhomogeneity
of the turbulent plasma, it bears a strong resemblance to the stochastic transition radiation \cite{KL08a,KL08b}. Let also note a possible
relation of this instability to that in nonrelativistic QED plasma discussed in \cite{Tsit}.

\subsubsection{Turbulent corrections for plasmons}

It is also of interest to analyze the effects of turbulence on the properties of collective excitations of QED plasma, the plasmons. The plasmons
are characterized by dispersion relations $\omega_{\rm T(L)} (\absvec{k})$ that are read from the solutions of dispersion equation for the corresponding components of
dielectric permittivity, which are just a real part  of zeroes of inverse transverse and longitudinal wave propagators:
\begin{equation}
\begin{split}
&Re\left[ \left. {\mathbf k}^{2}\left(1-\dfrac{\Pi_{\rm L}(k^{0},\absvec{k})}{\omega^{2}}\right)\right|_{k^{0}=\omega_{\rm L}(\absvec{k})}\right]=0\\
&Re\left[{\mathbf k}^{2}-(k^{0})^{2}+\Pi_{\rm T}((k^{0},\absvec{k})\mid_{k^{0}=\omega_{\rm T}(\absvec{k})}\right]=0
\end{split}
\end{equation}
Thus, real part of polarization tensor corresponds to propagation of plasmons in a medium, while it's imaginary part defies plasmon smearing.

Let us focus first on a shift of plasmons dispersion relations in turbulent medium. In general dispersion equations can be solved only numerically. Analytical expressions can be obtained in certain limits. Let us focus on the
deeply timelike regime of $x \gg 1$.  In non-turbulent HTL Vlasov plasma the time-like plasmon modes do not decay, since imaginary part of
polarization tensor in that limit is zero. For polarization tensor of the form (\ref{SilinPolTens}) and frequencies $\frac{k}{\omega_{pl}}<<1$ the
corresponding solutions of dispersion equations may be expanded as powers of $\frac{\absvec{k}}{\omega_{\rm pl}}$:
\begin{equation}
\begin{split}
 & \omega_{\rm L}^{2}(\absvec{k})_{\rm HTL}=\omega_{\rm pl}^{2}\left(1+\dfrac{3}{5}\left(\dfrac{\absvec{k}}{\omega_{\rm pl}}\right)^{2}+
 O\left(\left(\dfrac{\absvec{k}}{\omega_{\rm pl}}\right)^{4}\right)\right)\\
& \omega_{\rm T}^{2}(\absvec{k})_{\rm HTL} =\omega_{\rm pl}^{2} \left(1+\dfrac{6}{5}\left(\dfrac{\absvec{k}}{\omega_{\rm
pl}}\right)^{2}+O\left(\left(\dfrac{\absvec{k}}{\omega_{\rm pl}}\right)^{4}\right)\right)
\end{split}
\label{disprelHTL}
\end{equation}
where we have used a standard definition for the plasma frequency $\omega^2_{\rm pl}=m^2_D/3$.

In a turbulent plasma plasmons decay even in a Vlasov limit since polarization tensor has imaginary part. As to the
turbulent modifications of the HTL dispersion relation (\ref{disprelHTL}), it can be conveniently written as
\begin{equation}
\begin{split}
 &\omega_{\rm L}^2(\absvec{k})_{\rm turb}=(\omega^{\rm turb}_{\rm pl\; L})^2
 \left(1+\dfrac{3}{5}y_{\rm L}^2\right)- \dfrac{e^{4} l^{2}}{6 \pi^{2}}\left(\dfrac{24}{5}\langle E^{2}\rangle+\dfrac{64}{15}\langle B^{2}\rangle \right)y_{\rm L}^2+O\left(y_{\rm L}^4 \right)\\
 &\omega_{\rm T}^2(\absvec{k})_{\rm turb}=(\omega^{\rm turb}_{\rm pl\;T})^2\left(1+\dfrac{3}{5}y_{\rm T}^2\right)-\dfrac{e^{4} l^{2}}{6 \pi^{2}} \left( \dfrac{24}{7}\langle
E^{2}\rangle+\dfrac{32}{15}\langle B^{2}\rangle\ \right) y_{\rm T}^2+O\left(y_{\rm T}^4 \right)\; ,
\end{split}
\end{equation}
where
\begin{equation}
y_{\rm L}=\frac{\absvec{k}}{\omega^{\rm turb}_{\rm pl\;L}}; \;\;\;\; y_{\rm T}=\frac{\absvec{k}}{\omega^{\rm turb}_{\rm pl\;T}}\;,
\end{equation}
and
\begin{equation}
\begin{split}
&(\omega^{\rm turb}_{\rm pl\;L})^{2}=\omega_{\rm pl\;L }^{2}-\dfrac{e^{4} l^{2}}{6 \pi^{2}} \left(\dfrac{16}{3}\langle E^{2}\rangle+\dfrac{8}{3}\langle B^{2}\rangle\right)\\
&(\omega^{\rm turb}_{\rm pl\;T})^{2}=\omega_{\rm pl\;T }^{2}-\dfrac{e^{4} l^{2}}{6 \pi^{2}} \left(\dfrac{128}{15}\langle
E^{2}\rangle+\dfrac{8}{3}\langle B^{2}\rangle\right).
\end{split}
\end{equation}

Now let us consider plasmons smearing. As it can be easily seen that a rate of decay for plasmons is connected to an imaginary part of polarization tensor by a formula:
\begin{equation}
\Gamma_{T(L)}=\sqrt{-Im(\Pi_{T(L)})}
\end{equation}
In a timeline region considered above imaginary part of both transverse and longitudinal components of polarization tensor are lesser than zero: there i no instability for timeline modes. Also it should be noted that turbulent smearing is a leading order effect on $(k l)$ compared with a turbulent modification plasmon dispersion relations.

\section{Conclusions}

Let us briefly summarize the results obtained. In the present paper we have calculated the polarization properties of turbulent ultrarelativistic
QED plasma in the first order in the regular field and second order in the turbulent field. The main results are:
\begin{itemize}
\item The nonlinear Landau damping originating from turbulent effects corresponds to instability for the spacelike modes. At strong enough
turbulent fields this leads to an overall turbulent instability of a system.
\item Turbulent modifications of the plasmon dispersion relations were calculated.
\end{itemize}

\begin{center}
{\bf Acknowledgements}
\end{center}

M.K. and A.L. are grateful to I.M. Dremin and K.P. Zybin for useful discussions. Their work was partially supported by the RFBR grant 12-02-91504
(M.K. and A.L.) and RAS-CERN program (A.L.).

\section*{Appendix}

\appendix

In this Appendix we provide some details on calculation of the turbulent contribution $\langle \delta f_{12} \rangle_{\rm I}$ in
(\ref{AnsDiagr1}).

\section{Usability conditions}\label{Dupree}
System of equations (\ref{expcomp}) has been written up in an assumption that turbulent corrections don't modify equilibrium distribution $f^{\rm eq}=f_{00}$ in absence of regular field. Indeed, in a last line of (\ref{expcomp}) initial distribution $f^{\rm eq}$ is used instead of its  turbulent modification $f^{\rm 02}$.

To justify this assumption let us examine turbulent modification of initial condition. It can be easily seen that in a first order of turbulent pulsations, equation for distribution function takes form (this function is denoted as mean turbulent state, $f^{\rm m. t. s.}$):
\begin{equation}
(p^{\mu} \partial_{\mu}- p_{\mu} \langle F_{0 1}^{\mu \nu} \partial_{p \, \nu } p_{\mu^{\prime}} G  F_{0 1}^{\mu^{\prime}\nu^{\prime}} \rangle \partial_{p \, \nu^{\prime}} )f^{\rm m.t.s.}=0
\end{equation}
Further notation $\hat{O}(p,0,K)\equiv  p_{\mu} \langle F_{0 1}^{\mu \nu} \partial_{p \, \nu } p_{\mu^{\prime}} G  F_{0 1}^{\mu^{\prime}\nu^{\prime}} \rangle \partial_{p \, \nu^{\prime}}$ is used (here $K$ characterizes correlation of turbulent fields).  Let us examine whether this equation has stationary solutions. Stationary solution should be of a form:
\begin{equation}
f^{(R) 0}(k,p)=\delta^{4}(k)h(p)
\label{init_turb_state_3}
\end{equation}

Thus, $h(p)$ should be a solution of equation:
\begin{equation}
\hat{O}(p,0,K)h(p)=0
\end{equation}
However, solution of this equation in isotropic turbulent plasma (in a first order of turbulent pulsations) is power like:
\begin{equation}
h(p)=\dfrac{C_{1}}{p}
\end{equation}

Another option is to find non-stationary solutions of (\ref{init_turb_state_2}). Simplest non-stationary solution that approach to $f^{eq}(p)\delta^{4}(k)$ if $K\rightarrow 0$ has the form:
\begin{equation}
f^{(R) 0}(k,p)=\imath p^{0}\delta (\imath(pk)-\hat{O}(p,k,K))\delta^{3}(k)f^{eq}(p)
\end{equation}
In spatial representation it takes:
\begin{equation}
f^{(R) 0}(x,p)=\exp\left(\hat{\kappa} t \right)f^{eq}(p)
\end{equation}
here $\hat{\kappa}$ is a solution of equation:
\begin{equation}
\imath(p k^{0})-\hat{O}(p,k^{0},0,K)=0
\end{equation}
(notice that $\hat{\kappa}$ is an operator itself)

In a linear response theory of non-stationar matter polarization operator is non-local, polarization depends not only on regular field $F_{\mu \nu}^{R}(k)$ but also on its derivative $\partial_{k}F_{\mu \nu}^{R}(k)$  In diagrammatic notations non-stationary contribution to polarization comes from a simplest diagram (the same that gives first order HTL answer):
\begin{fmffile}{diag_23_2}
\begin{equation}
\parbox{20mm}{
\begin{fmfgraph*}(20,15)
\fmfleft{i1,i2,i3} \fmfright{o1,o2,o3}
\fmf{phantom}{i1,o1}
\fmf{dashes}{i2,v2}
\fmf{plain,lab.dist=.05w,lab.side=right,lab=$R$}{v2,o2}
\fmf{phantom,tension=+0.3}{i3,v3}
\fmf{phantom}{v3,o3}
\fmf{dbl_wiggly,lab.dist=.05w,lab.side=left,lab=$R$,tension=-0.01}{v2,v3}
\fmfdot{v2}
\end{fmfgraph*}}
\end{equation}
\end{fmffile}

Writing this diagram explicitly one gets:
\begin{equation}
\begin{split}
&\dfrac{1}{\imath((pk)+\imath\epsilon)}p_{\mu}\dfrac{\partial}{\partial p^{\nu}}\int d^{4} k_{1} e q F^{\mu \nu}(k_{1}) \imath p^{0} \delta(\imath(p (k-k_{1}))-\hat{O}(p,k-k_{1},K))\delta^{3}(\mathbf{k}) f^{eq}\\
&=\dfrac{e q}{\imath((pk)+\imath\epsilon)}p_{\mu}\dfrac{\partial}{\partial p^{\nu}}F^{\mu \nu}\left(\hat{\kappa},\mathbf{k}\right)f^{eq}(p)=\dfrac{e  q}{\imath((pk)+\imath\epsilon)}p_{\mu}\dfrac{\partial}{\partial p^{\nu}}\left(F^{\mu \nu}(k)+\imath\dfrac{\partial F^{\mu \nu}(k)}{\partial k^{0}}\dfrac{\hat{O}(p,0,K)}{p^{0}}\right)f^{eq}(p)
\end{split}
\end{equation}
term in a last line may be neglected in comparison with a diagram (see below)
\begin{fmffile}{diag20_1}
\begin{equation}
\parbox{35mm}{\begin{fmfgraph*}(35,15)
\fmfleft{i1,i2,i3} \fmfright{o1,o2,o3}
\fmf{phantom}{i1,o1}
\fmf{dashes}{i2,v1}
\fmf{dashes}{v1,v2}
\fmf{dashes}{v2,v3}
\fmf{plain,lab.dist=.05w,lab.side=right,lab=$eq$}{v3,o2}
\fmf{phantom,tension=1.4}{i3,u1}
\fmf{phantom,tension=1.5}{u1,u2}
\fmf{phantom,tension=1.6}{u2,u3}
\fmf{phantom,tension=9.}{u3,o3}
\fmffreeze
\fmf{wiggly,lab.dist=.05w,lab.side=left,lab=$K$,left,tension=-0.5}{v1,v2}
\fmf{dbl_wiggly,lab.dist=.05w,lab.side=left,lab=$R$,tension=-0.01}{v3,u3}
\fmfdot{v1}
\fmfdot{v2}
\fmfdot{v3}
\end{fmfgraph*}}
\end{equation}
\end{fmffile}
if:
\begin{equation}
\dfrac{{\partial F_{\mu \nu}^{R}(k)/\partial k^{0}}}{F_{\mu \nu}^{R}(k)}<<\cfrac{1}{k}
\end{equation}
in spatial representation this transforms to:
\begin{equation}
t F(t) << \int_{0}^{t} d \tau |F(\tau)|
\end{equation}
which means that fields should change in time slow enough

Another limitation comes from an applicability of perturbative expansion on turbulent pulsations. Later works if turbulent
contributions should remain smaller than linear HTL contribution:

\begin{equation}
\delta f_{12} << \delta f_{HTL}
\end{equation}
If later does not hold loops with more than one turbulent correlations of a type:\\
\begin{fmffile}{diagabel4_1_2}
\begin{tabular}{lll}
\parbox{0.4\linewidth}{\begin{center}\begin{fmfgraph*}(15,25)
\fmfleft{i1} \fmfright{o1}
\fmf{dashes}{i1,i2}
\fmf{dashes}{i2,v}
\fmf{dashes}{v,o2}
\fmf{dashes}{o2,o1}
\fmffreeze
\fmf{wiggly,lab.side=left,lab=$K$,left}{i1,o1}
\fmf{wiggly,right}{i2,o2}
\fmfdot{i1,i2,o1,o2}
\end{fmfgraph*}
\end{center}}& \parbox{0.2\linewidth}{and}&
\parbox{0.4\linewidth}{\begin{fmfgraph*}(15,25)
\fmfleft{i1} \fmfright{o1}
\fmf{dashes}{i1,i2}
\fmf{dashes}{i2,v}
\fmf{dashes}{v,o2}
\fmf{dashes}{o2,o1}
\fmffreeze
\fmf{wiggly,left}{i1,o2}
\fmf{wiggly,right}{i2,o1}
\fmfdot{i1,i2,o1,o2}
\end{fmfgraph*}}
\end{tabular}
\end{fmffile}
start to contribute. Thus, limit holds until contribution to polarization in plasma is smaller than first order HTL contribution. As it is shown below, in case of purely magnetic turbulent fields for example this happens for field correlations of order:
\begin{equation}
\dfrac{4}{\pi^{2}\sqrt{\pi}}\dfrac{(\absvec{k}l)}{\mathbf{k}^2} \dfrac{e^2\langle B^2 \rangle}{T^2 } \Phi(x) \sim 1
\end{equation}
where $\Phi$ is a function of $x = \dfrac{\omega}{|\mathbf{k}|}$ defined below.

Also applicability conditions of linear response theory give restriction to amplitude of a regular field. Latter may be used when regular field is not strong enough to significantly change particle distribution in plasma Following simple physical arguments \cite{AMY2005} may be used: In abelian plasma regular field may significantly change distribution of plasma particles if momentum that particle with a typical momentum in plasma receives from a field on its wavelength is compatible with a particle momentum itself (which is of order $T$). This gives condition:
\begin{equation}
| F_{\mu \nu}|<<\dfrac{k_{field} T}{e}
\end{equation}

Another restriction comes from a loop integration. Integrals are performed in a limit $k l<<1$ where $l$ is a correlations. Also supposing additional condition $k<<g T$ as it may be shown in this limit  there is no contribution to $f_{1 2}$ coming from a terms with $F_{11}$ and $F_{12}$

\section{Loop integration over $k_1$}

Let us first perform loop integrations over $k_1$ in the expressions (\ref{AnsDiagr1}). We will use the following expression for the correlator of
the turbulent fields Let us stress that although in the main body of the paper we display only answers for the case of isotropic turbulence, in
the calculations described in this Appendix we assume a generic tensorial structure of the correlator $K^{\mu \nu \mu^{\prime} \nu^{\prime}}(x)$.
We have to calculate the following integral:
\begin{equation}\label{intI1}
I_1 = \int d^{4} k_{1} \dfrac{K^{\mu \nu \mu^{\prime}\nu^{\prime}}(k_{1})}{\imath((p (k-k_{1}))+\imath \epsilon)}
\end{equation}
Using the $\alpha$-representation
\begin{equation}
\dfrac{1}{\imath((p (k-k_{1}))+\imath \epsilon)}=-\int_{0}^{+\infty}d \alpha \exp\left[\imath\alpha((p (k-k_{1}))+\imath \epsilon)\right]
\end{equation}
and the explicit parametrization of the turbulent correlator (\ref{turbcor}) we get
\begin{equation}
\begin{split}
&I_1=-\dfrac{a^{3}\tau}{(2\pi)^{2}}K_{0}^{\mu \nu \mu^{\prime}\nu^{\prime}}\int_{0}^{+\infty}d \alpha \int d^{4} k_{1}\exp\left[-\dfrac{\mathbf{k}_{1}^{2}a^{2}}{2}-\dfrac{\left(k_{1}^{0}\right)^{2}\tau^{2}}{2}\right]\exp\left[\imath \alpha \left(p(k-k_{1})+\imath \epsilon \right)\right]\\
&=-\dfrac{a^{3}\tau}{(2\pi)^{2}}K_{0}^{\mu \nu \mu^{\prime} \nu^{\prime}}\int_{0}^{+\infty}d \alpha \int d^{4} k_{1}\exp\left[-\dfrac{a^{2}}{2}\left(\mathbf{k}_{1}^{2}-\dfrac{2\imath \alpha \mathbf{p k_{1}}}{a^{2}}\right)-\dfrac{\tau^{2}}{2}\left((k_{1}^{0})^{2}+\dfrac{2\imath\alpha p^{0}k^{0}}{\tau^{2}}\right)\right]\times\\
&\times\exp[\imath \alpha (p k)]=-K_{0}^{\mu \nu \mu^{\prime}\nu^{\prime}}\int_{0}^{+\infty} d \alpha \exp\left[-\dfrac{\alpha^{2}\mathbf{p}^{2}}{2 a^{2}}-\dfrac{\alpha^{2}(p^{0})^{2}}{2 \tau^{2}}+\imath \alpha (p k)\right]=\\
&\left(c.o.v. \; \hat{\alpha}=-\alpha\cdot\sqrt{\frac{\mathbf{p}^{2}}{ 2 a^{2}}+\frac{\left( p^{0}\right)^{2}}{2\tau^{2}}} \right)=-K_{0}^{\mu \nu \mu^{\prime}\nu^{\prime}}\dfrac{1}{\sqrt{\frac{\mathbf{p}^{2}}{2 a^{2}}+\frac{\left(p^{0}\right)^{2}}{2 \tau^{2}}}}\int_{0}^{+\infty}d \hat{\alpha}\exp\left[-\hat{\alpha}^{2}+\imath\hat{\alpha}(p k)\left/\sqrt{\frac{\mathbf{p}^{2}}{2 a^{2}}+\frac{\left(p^{0}\right)^{2}}{2 \tau^{2}}}\right.\right]\approx\\
&\approx - K_{0}^{\mu \nu \mu^{\prime}}\dfrac{1}{\sqrt{\frac{\mathbf{p}^{2}}{2 a^{2}}+\frac{\left(p^{0}\right)^{2}}{2\tau^{2}}}}\int_{0}^{+\infty}d \hat{\alpha}\left[1+\dfrac{\imath \hat{\alpha} (pk)}{\sqrt{\frac{\mathbf{p}^{2}}{2a^{2}}+\frac{\left(p^{0}\right)^{2}}{2\tau^{2}}}} \right]\exp[-\hat{\alpha}^{2}]
\end{split}
\end{equation}
We have
\begin{equation}
I_1=-K_{0}^{\mu\nu\mu^{\prime}\nu^{\prime}}\dfrac{1}{\sqrt{\frac{\mathbf{p}^{2}}{2 a^{2}}+\frac{\left(p^{0}\right)^{2}}{2\tau^{2}}}}\left[\dfrac{\sqrt{\pi}}{2}+\dfrac{\imath (pk)}{2\sqrt{\frac{\mathbf{p}^{2}}{2a^{2}}+\frac{\left(p^{0}\right)^{2}}{2\tau^{2}}}}\right]
\end{equation}
and, finally, get for the turbulent contribution $\langle \delta f_{12} \rangle_{\rm I}$ in  (\ref{AnsDiagr1}):
\begin{fmffile}{diag20_1}
\begin{equation}
\begin{split}
\langle \delta f_{12} \rangle_{\rm I} =
\parbox{35mm}{\begin{fmfgraph*}(35,15)
\fmfleft{i1,i2,i3} \fmfright{o1,o2,o3}
\fmf{phantom}{i1,o1}
\fmf{dashes}{i2,v1}
\fmf{dashes}{v1,v2}
\fmf{dashes}{v2,v3}
\fmf{plain,lab.dist=.05w,lab.side=right,lab=$eq$}{v3,o2}
\fmf{phantom,tension=1.4}{i3,u1}
\fmf{phantom,tension=1.5}{u1,u2}
\fmf{phantom,tension=1.6}{u2,u3}
\fmf{phantom,tension=9.}{u3,o3}
\fmffreeze
\fmf{wiggly,lab.dist=.05w,lab.side=left,lab=$K$,left,tension=-0.5}{v1,v2}
\fmf{dbl_wiggly,lab.dist=.05w,lab.side=left,lab=$R$,tension=-0.01}{v3,u3}
\fmfdot{v1}
\fmfdot{v2}
\fmfdot{v3}
\end{fmfgraph*}}\approx&-\dfrac{e^{3}\,q^{3}}{\imath ((p k)+
\imath \epsilon)}F^{\rho \sigma}(k)K_{0}^{\mu \nu \mu^{\prime} \nu^{\prime}} \\ p_{\mu}\dfrac{\partial}{\partial p^{\nu}}
\left[\dfrac{\sqrt{\pi}}{2\sqrt{\frac{\mathbf{p}^{2}}{2 a^{2}}+\frac{\left(p^{0}\right)^{2}}{2\tau^{2}}}}+ \dfrac{\imath
(pk)}{2\left(\frac{\mathbf{p}^{2}}{2 a^{2}}+\frac{\left(p^{0}\right)^{2}}{2\tau^{2}}\right)}\right] &\times
p_{\mu^{\prime}}\dfrac{\partial}{\partial p^{\nu^{\prime}}}\dfrac{1}{\imath ((p k)+\imath \epsilon)}p_{\rho}\dfrac{\partial}{\partial
p^{\sigma}}f_{\rm F}(p,T)
\end{split}
\end{equation}
\end{fmffile}
where $f_{\rm F}(p,T)=f_{\rm F}(p^0,T)$ is a thermal Fermi distribution.

\section{Contribution to the induced current}

Let us turn to the computation of the induced current corresponding to the diagram (\ref{Diagr1}). To the leading order in the inhomogeneity this is the only contribution we have to keep. We have
\begin{equation}
\begin{split}
\delta j_{\; \rm I}^\lambda (k)&=e \sum_{q,s} q \int d^{4}p \,\delta(p^{2})\theta(p^{0}) p^{\lambda} \langle \delta f_{12} (p,k,q) \rangle_{\rm I}=
-e^4 K_{0}^{\mu\nu\mu^{\prime}\nu^{\prime}}F^{\rho \sigma}(k) \sum_{q,s} q^4\\
&\times \int d^{4}p \,\delta(p^{2})\theta(p^{0}) p^{\lambda} \dfrac{p_{\mu}}{\imath ((p k)+\imath \epsilon)}\dfrac{\partial}{\partial p^{\nu}}\left[\dfrac{\sqrt{\pi}}{2\sqrt{\frac{\mathbf{p}^{2}}{2 a^{2}}+\frac{\left(p^{0}\right)^{2}}{2\tau^{2}}}}+\dfrac{\imath (pk)}{2\left(\frac{\mathbf{p}^{2}}{2 a^{2}}+\frac{\left(p^{0}\right)^{2}}{2\tau^{2}}\right)}\right] p_{\mu^{\prime}}\dfrac{\partial}{\partial p^{\nu^{\prime}}}\\
&\dfrac{1}{\imath ((p k)+\imath \epsilon)}p_{\rho}\dfrac{\partial}{\partial p^{\sigma}}f_{\rm F}(p,T)
\end{split}
\end{equation}
Integrating by parts and taking into account the asymmetry of $K^{\mu \nu \mu^{\prime}\nu^{\prime}}$ in $\mu\, \nu$ and $\mu^{\prime}\,
\nu^{\prime}$ one gets:
\begin{equation}\label{j11}
\begin{split}
\delta j_{\; \rm I}^\lambda (k) &=-e^{4}K_{0}^{\mu\nu\mu^{\prime}\nu^{\prime}}F^{\rho \sigma}(k)\sum_{q,s} q^4\int d^{4}p \delta^{4}(p^{2})
\theta(p^{0})
\left[\dfrac{\delta_{\nu}^{\lambda}}{(p k)}-\dfrac{k_{\nu}p^{\lambda}}{(p k)^{2}} \right]p_{\mu}\\
&\left[\dfrac{\sqrt{\pi}}{2\sqrt{\frac{\mathbf{p}^{2}}{2 a^{2}}+\frac{\left(p^{0}\right)^{2}}{2\tau^{2}}}}+\dfrac{\imath
(pk)}{2\left(\frac{\mathbf{p}^{2}}{2 a^{2}}+ \frac{\left(p^{0}\right)^{2}}{2\tau^{2}}\right)}\right]p_{\mu^{\prime}}
\left[\vphantom{\dfrac{\sqrt{\pi}}{2\sqrt{\frac{\mathbf{p}^{2}}{2 a^{2}}+
\frac{\left(p^{0}\right)^{2}}{2\tau^{2}}}}}\dfrac{\partial^{2}f_F}{\partial p^{\nu^{\prime}}\partial p^{\sigma}}\dfrac{p_{\rho}}{(pk)}+
\dfrac{\partial f_F}{\partial p^{\sigma}}\dfrac{g_{\rho \nu^{\prime}}}{(pk)}-\dfrac{k_{\nu^{\prime}}p_{\rho}}{(pk)^{2}}\dfrac{\partial f_F}{\partial p^{\sigma}}\right]\\
& \equiv P_{11}^{\lambda}+P_{12}^{\lambda}+P_{21}^{\lambda}+P_{22}^{\lambda}+P_{31}^{\lambda}+P_{32}^{\lambda}
+P_{41}^{\lambda}+P_{42}^{\lambda}+P_{51}^{\lambda}+P_{52}^{\lambda}+P_{61}^{\lambda}+P_{62}^{\lambda}
\end{split}
\end{equation}
Let us note that the gauge invariance of the polarization operator (its transversality with respect to $k_\mu$) is obvious from the structure of
the first square brackets in (\ref{j11}).

\subsection{Useful notations}

Let us introduce some useful notations and definitions that help to present the explicit expressions for (\ref{j11}) in a readable form.

In calculating the integrals over spatial momenta it is convenient to use the following notations:

\begin{equation}
\hat{\delta}_{\mu\nu} = \rm{diag} (0,1,1,1); \;\;\;\;\; \hat{g}_{\mu\nu} = \rm{diag} (0,-1,-1,-1); \;\;\;\;\; \hat{k}_{\mu}=(0,-\mathbf{k}) ;
\;\;\;\;\;\; \hat{e}_\mu = \frac{\hat{k}_{\mu}}{\mathbf{k}}
\end{equation}

\begin{equation}
 {\cal P}_{\mu \nu}^{\rm L} (\hat{k})= \dfrac{\hat{k}_\mu \hat{k}_\nu}{\mathbf{k}^2}; \;\;\;\;
 {\cal P}_{\mu \nu}^{\rm T} (\hat{k})= \left( -\hat{g}_{\mu \nu}-\dfrac{\hat{k}_\mu \hat{k}_\nu}{\mathbf{k}^2} \right)
\end{equation}

The the expressions for the standard integrals arising after angular integration $A_{1 \cdot 9}$, $B_{1 \cdot 9}$, $C_{1 \cdot 9}$ and $D_{1 \cdot
9}$ are listed below in \ref{standint}.

\subsection{Calculating the integrals: examples}

Let us illustrate the calculations at the examples of the first two contributions $P_{11}^\lambda$ and $P_{12}^\lambda$ in (\ref{j11}). We have
\begin{equation}
\begin{split}
 P_{11}^{\lambda} &=-\dfrac{ 4\cdot 2 e^{4}\sqrt{\pi} l K_{0}^{\mu\nu\mu^{\prime}\nu^{\prime}}F^{\rho \sigma}}{2 (2\pi )^{3}}
 \left(\int_{0}^{+\infty}d p p \dfrac{\partial^{2} f}{\partial p^{2}}\right)\int \Omega_{\mathbf{v}} \dfrac{g_{\nu^{\prime} 0}
 g_{\sigma 0}\delta_{\nu}^{\lambda}v_{\mu}v_{\rho}v_{\mu^{\prime}}}{(v k)^{2}}=\\
 &=\dfrac{2 \imath \sqrt{\pi} g^{4} l}{(\pi)^{2}} \langle F_{0}^{\mu \lambda}F_{0}^{\mu^{\prime} 0}\rangle\left(k^{\rho} A^{\sigma}
 -k^{\sigma}A^{\rho}\right)g_{\sigma 0}
 \left[g_{\mu 0}\cdot {\cal P}^{\rm L}_{\mu^\prime \rho}(\hat{k})\cdot B_{3}(\omega,{\absvec{k}})+
      g_{\mu 0} \cdot {\cal P}^{\rm T}_{\mu^\prime \rho}(\hat{k}) \cdot B_{4} (\omega,{\absvec{k}}) \right.\\
 &\left. +\hat{e}_{\mu^\prime}\hat{e}_\rho \hat{e}_\mu \cdot B_{5}(\omega,{\absvec{k}})+
  \left( \hat{e}_{\mu^\prime} \cdot {\cal P}^{\rm T}_{\rho \mu}(\hat{k}) +
         \hat{e}_{\mu} \cdot {\cal P}^{\rm T}_{\rho \mu^\prime}(\hat{k}) +
          \hat{e}_{\rho} \cdot {\cal P}^{\rm T}_{\mu^\prime \mu}(\hat{k}) \right)B_{6}(\omega,{\absvec{k}}) \right]
\end{split}
\end{equation}
\begin{equation}
\begin{split}
 &P^{\lambda}_{12}=\dfrac{ 4\cdot 2 e^{4}l^{2} K_{0}^{\mu\nu\mu^{\prime}\nu^{\prime}}F^{\rho \sigma}}{2 (2\pi )^{3}}
 \left(\int_{0}^{+\infty}d p p \dfrac{\partial^{2} f}{\partial p^{2}}\right)
 \int \Omega_{\mathbf{v}} \dfrac{g_{\nu^{\prime} 0} g_{\sigma 0}\delta_{\nu}^{\lambda}v_{\mu}v_{\rho}v_{\mu^{\prime}}}{(v k)}=\\
 &=-\dfrac{2 \imath e^{4} l^{2}}{(\pi)^{2}} \langle F_{0}^{\mu \lambda}F_{0}^{\mu^{\prime} 0}\rangle\left(k^{\rho} A^{\sigma}-k^{\sigma}A^{\rho}\right)g_{\sigma 0}
 \left[g_{\mu 0}\cdot {\cal P}^{\rm L}_{\mu^\prime \rho}(\hat{k})\cdot A_{3}(\omega,{\absvec{k}})+
      g_{\mu 0} \cdot {\cal P}^{\rm T}_{\mu^\prime \rho}(\hat{k}) \cdot A_{4} (\omega,{\absvec{k}}) \right.\\
 &\left. +\hat{e}_{\mu^\prime}\hat{e}_\rho \hat{e}_\mu \cdot A_{5}(\omega,{\absvec{k}})+
  \left( \hat{e}_{\mu^\prime} \cdot {\cal P}^{\rm T}_{\rho \mu}(\hat{k}) +
         \hat{e}_{\mu} \cdot {\cal P}^{\rm T}_{\rho \mu^\prime}(\hat{k}) +
          \hat{e}_{\rho} \cdot {\cal P}^{\rm T}_{\mu^\prime \mu}(\hat{k}) \right)A_{6}(\omega,{\absvec{k}}) \right]
\end{split}
\end{equation}

\subsection{Calculating the integrals: answers}

\begin{equation}
\begin{split}
 &P^{\lambda}_{21}=\dfrac{2 \imath \sqrt{\pi} e^{4} l}{(\pi)^{2}} \langle F_{0}^{\mu \nu}F_{0}^{\mu^{\prime} 0}\rangle\left(k^{\rho} A^{\sigma}-k^{\sigma}A^{\rho}\right)g_{\sigma 0} k_{\nu}\left[g_{\mu 0}g^{\lambda 0}\cdot {\cal P}^{\rm L}_{\mu^{\prime} \rho}(\hat{k}) C_{3}(\omega,{\absvec{k}})+g_{\mu 0}g^{\lambda 0} \cdot {\cal P}^{\rm T}_{\mu^{\prime} \rho}(\hat{k}) C_{4}(\omega,{\absvec{k}})\right.\\
&\left.+ \left(g_{\mu 0}\hat{e}_{\mu^\prime}\hat{e}_\rho \hat{e}^\lambda+ g^{\lambda 0} \hat{e}_{\mu^\prime}\hat{e}_\rho \hat{e}_\mu\right)C_{5}(\omega,{\absvec{k}})+\left(g_{\mu 0}\left[\hat{e}_\rho\cdot{\cal P}^{{\rm T} \lambda}_{\mu^{\prime}}(\hat{k})+\hat{e}_{\mu^{\prime}}\cdot{\cal P}^{{\rm T} \lambda}_{\rho}(\hat{k})+\hat{e}^{\lambda}\cdot{\cal P}^{\rm T}_{\mu^{\prime} \rho}(\hat{k})\right]+\right.\right.\\
&\left.\left.g^{\lambda 0}\left[\hat{e}_{\rho}\cdot{\cal P}^{\rm T}_{\mu^{\prime} \mu}(\hat{k})+\hat{e}_{\mu^{\prime}}\cdot{\cal P}^{\rm T}_{\rho \mu}(\hat{k})+\hat{e}_{\mu}\cdot{\cal P}^{\rm T}_{\rho \mu^{\prime}}(\hat{k}) \right] \right)C_{6}(\omega,{\absvec{k}})+\left[\hat{e}_{\mu}\hat{e}_{\mu^{\prime}}\cdot{\cal P}^{{\rm T} \lambda}_{\rho}(\hat{k})+\hat{e}_{\mu}\hat{e}_{\rho}\cdot{\cal P}^{{\rm T} \lambda}_{\mu^{\prime}}(\hat{k})+ \right.\right.\\
&\left.\left. \hat{e}_{\mu^{\prime}}\hat{e}_{\rho}\cdot{\cal P}^{{\rm T} \lambda}_{\mu}(\hat{k}) + \hat{e}_{\mu}\hat{e}^{\lambda}\cdot{\cal P}^{\rm T}_{\mu^{\prime} \rho}(\hat{k}) + \hat{e}_{\mu^{\prime}}\hat{e}^{\lambda}\cdot{\cal P}^{\rm T}_{\mu \rho}(\hat{k}) + \hat{e}_{\rho}\hat{e}^{\lambda}\cdot{\cal P}^{\rm T}_{\mu \mu^{\prime}}(\hat{k})   \right]C_{8}(\omega,{\absvec{k}})+\hat{e}_{\mu^{\prime}}\hat{e}_{\mu}\hat{e}_{\rho}\hat{e}^{\lambda}C_{7}(\omega,{\absvec{k}})\right.\\
&\left.+\left({\cal P}^{{\rm T} \lambda}_{\mu^{\prime}}(\hat{k})\cdot{\cal P}^{{\rm T}}_{\mu \rho}(\hat{k})+{\cal P}^{{\rm T}
\lambda}_{\mu}(\hat{k})\cdot{\cal P}^{{\rm T}}_{\mu^{\prime} \rho}(\hat{k})+{\cal P}^{{\rm T} \lambda}_{\rho}(\hat{k})\cdot{\cal P}^{{\rm
T}}_{\mu^{\prime} \mu}(\hat{k})\right)C_{9}(\omega,{\absvec{k}}) \right]
\end{split}
\end{equation}

\begin{equation}
\begin{split}
&P^{\lambda}_{22}=-\dfrac{2 e^{4} l^{2}}{(\pi)^{2}} \langle F_{0}^{\mu \nu}F_{0}^{\mu^{\prime} 0}\rangle\left(k^{\rho} A^{\sigma}-k^{\sigma}A^{\rho}\right)g_{\sigma 0} k_{\nu}\left[g_{\mu 0}g^{\lambda 0}\cdot {\cal P}^{\rm L}_{\mu^{\prime} \rho}(\hat{k}) B_{3}(\omega,{\absvec{k}})+g_{\mu 0}g^{\lambda 0} \cdot {\cal P}^{\rm T}_{\mu^{\prime} \rho}(\hat{k}) B_{4}(\omega,{\absvec{k}}) \right.\\
&\left.+ \left(g_{\mu 0}\hat{e}_{\mu^\prime}\hat{e}_\rho \hat{e}^\lambda+ g^{\lambda 0} \hat{e}_{\mu^\prime}\hat{e}_\rho \hat{e}_\mu\right)B_{5}(\omega,{\absvec{k}})+\left(g_{\mu 0}\left[\hat{e}_\rho\cdot{\cal P}^{{\rm T} \lambda}_{\mu^{\prime}}(\hat{k})+\hat{e}_{\mu^{\prime}}\cdot{\cal P}^{{\rm T} \lambda}_{\rho}(\hat{k})+\hat{e}^{\lambda}\cdot{\cal P}^{\rm T}_{\mu^{\prime} \rho}(\hat{k})\right]+\right.\right.\\
&\left.\left.g^{\lambda 0}\left[\hat{e}_{\rho}\cdot{\cal P}^{\rm T}_{\mu^{\prime} \mu}(\hat{k})+\hat{e}_{\mu^{\prime}}\cdot{\cal P}^{\rm T}_{\rho \mu}(\hat{k})+\hat{e}_{\mu}\cdot{\cal P}^{\rm T}_{\rho \mu^{\prime}}(\hat{k}) \right] \right)B_{6}(\omega,{\absvec{k}})+\left[\hat{e}_{\mu}\hat{e}_{\mu^{\prime}}\cdot{\cal P}^{{\rm T} \lambda}_{\rho}(\hat{k})+\hat{e}_{\mu}\hat{e}_{\rho}\cdot{\cal P}^{{\rm T} \lambda}_{\mu^{\prime}}(\hat{k})+ \right.\right.\\
&\left.\left. \hat{e}_{\mu^{\prime}}\hat{e}_{\rho}\cdot{\cal P}^{{\rm T} \lambda}_{\mu}(\hat{k}) + \hat{e}_{\mu}\hat{e}^{\lambda}\cdot{\cal P}^{\rm T}_{\mu^{\prime} \rho}(\hat{k}) + \hat{e}_{\mu^{\prime}}\hat{e}^{\lambda}\cdot{\cal P}^{\rm T}_{\mu \rho}(\hat{k}) + \hat{e}_{\rho}\hat{e}^{\lambda}\cdot{\cal P}^{\rm T}_{\mu \mu^{\prime}}(\hat{k})   \right]B_{8}(\omega,{\absvec{k}})+\hat{e}_{\mu^{\prime}}\hat{e}_{\mu}\hat{e}_{\rho}\hat{e}^{\lambda}B_{7}(\omega,{\absvec{k}})\right.\\
&\left.+\left({\cal P}^{{\rm T} \lambda}_{\mu^{\prime}}(\hat{k})\cdot{\cal P}^{{\rm T}}_{\mu \rho}(\hat{k})+{\cal P}^{{\rm T}
\lambda}_{\mu}(\hat{k})\cdot{\cal P}^{{\rm T}}_{\mu^{\prime} \rho}(\hat{k})+{\cal P}^{{\rm T} \lambda}_{\rho}(\hat{k})\cdot{\cal P}^{{\rm
T}}_{\mu^{\prime} \mu}(\hat{k})\right)B_{9}(\omega,{\absvec{k}}) \right]
\end{split}
\end{equation}
\begin{equation}
\begin{split}
&P^{\lambda}_{31}=\dfrac{2 \imath\sqrt{\pi} e^{4} l}{(\pi)^{2}} \langle F_{0}^{\mu \nu}F_{0}^{\mu^{\prime} \nu^{\prime}}\rangle\left(k^{\rho} A^{\sigma}-k^{\sigma}A^{\rho}\right)g_{\sigma 0} g_{\rho \nu^{\prime}}\delta_{\nu}^{\lambda}\left[g_{\mu 0}g_{\mu^{\prime} 0}B_{1}(\omega,{\absvec{k}})+\left(g_{\mu 0}\hat{e}_{\mu^{\prime}}+g_{\mu^{\prime} 0}\hat{e}_{\mu}\right)B_{2}(\omega,{\absvec{k}})+\right.\\
&\left.{\cal P}^{{\rm L}}_{\mu^{\prime} \mu}(\hat{k})B_{3}(\omega,{\absvec{k}})+{\cal P}^{{\rm T}}_{\mu^{\prime}
\mu}(\hat{k})B_{4}(\omega,{\absvec{k}})\right]
\end{split}
\end{equation}
\begin{equation}
\begin{split}
&P^{\lambda}_{32}=-\dfrac{2 e^{4} l^{2}}{(\pi)^{2}} \langle F_{0}^{\mu \nu}F_{0}^{\mu^{\prime} \nu^{\prime}}\rangle\left(k^{\rho} A^{\sigma}-k^{\sigma}A^{\rho}\right)g_{\sigma 0} g_{\rho \nu^{\prime}}\delta_{\nu}^{\lambda}\left[g_{\mu 0}g_{\mu^{\prime} 0}A_{1}(\omega,{\absvec{k}})+\left(g_{\mu 0}\hat{e}_{\mu^{\prime}}+g_{\mu^{\prime} 0}\hat{e}_{\mu}\right)A_{2}(\omega,{\absvec{k}})+\right.\\
&\left.{\cal P}^{{\rm L}}_{\mu^{\prime} \mu}(\hat{k})A_{3}(\omega,{\absvec{k}})+{\cal P}^{{\rm T}}_{\mu^{\prime}
\mu}(\hat{k})A_{4}(\omega,{\absvec{k}})\right]
\end{split}
\end{equation}
\begin{equation}
\begin{split}
&P_{41}=-\dfrac{2 \imath\sqrt{\pi} e^{4} l}{(\pi)^{2}} \langle F_{0}^{\mu \nu}F_{0}^{\mu^{\prime} \nu^{\prime}}\rangle\left(k^{\rho} A^{\sigma}-k^{\sigma}A^{\rho}\right)g_{\sigma 0} g_{\rho \nu^{\prime}}k_{\nu}\left[g_{\mu^{\prime}0}g_{\mu 0}g^{\lambda 0}C_{1}(\omega,{\absvec{k}})+\left(g_{\mu^{\prime}0}g_{\mu 0}\cdot\hat{e}^{\lambda}+\right.\right.\\
&\left.\left.+g_{\mu 0}g^{\lambda 0}\cdot\hat{e}_{\mu^{\prime}}+g_{\mu^{\prime}0}g^{\lambda 0}\cdot\hat{e}_{\mu}\right)C_{2}(\omega,{\absvec{k}})+\left(g_{\mu^{\prime}0}\cdot{\cal P}^{{\rm L}\lambda}_{\mu}(\hat{k})+g_{\mu 0}\cdot{\cal P}^{{\rm L}\lambda}_{\mu^{\prime}}(\hat{k})+g^{\lambda 0}\cdot{\cal P}^{{\rm L}}_{\mu \mu^{\prime}}(\hat{k})\right)C_{3}(\omega,{\absvec{k}})+\right.\\
&\left.+\left[g_{\mu^{\prime} 0}\cdot{\cal P}^{{\rm T}\lambda}_{\mu}(\hat{k})+g_{\mu 0}\cdot{\cal P}^{{\rm T}\lambda}_{\mu^{\prime}}(\hat{k})+g^{\lambda 0}\cdot{\cal P}^{{\rm T}}_{\mu \mu^{\prime}}(\hat{k})\right]C_{4}(\omega,{\absvec{k}})+\hat{e}_{\mu^{\prime}}\hat{e}_{\mu}\hat{e}^{\lambda}C_{5}(\omega,{\absvec{k}})\right.\\
&\left.+\left[\hat{e}_{\mu}\cdot{\cal P}^{{\rm T}\lambda}_{\mu^{\prime}}(\hat{k})+\hat{e}_{\mu^{\prime}}\cdot{\cal P}^{{\rm
T}\lambda}_{\mu}(\hat{k})+\hat{e}^{\lambda}\cdot{\cal P}^{{\rm T}}_{\mu\mu^{\prime}}(\hat{k})\right]C_{6}(\omega,{\absvec{k}}) \right]
\end{split}
\end{equation}
\begin{equation}
\begin{split}
&P_{42}=\dfrac{2 e^{4} l^{2}}{(\pi)^{2}} \langle F_{0}^{\mu \nu}F_{0}^{\mu^{\prime} \nu^{\prime}}\rangle\left(k^{\rho} A^{\sigma}-k^{\sigma}A^{\rho}\right)g_{\sigma 0} g_{\rho \nu^{\prime}}k_{\nu}\left[g_{\mu^{\prime}0}g_{\mu 0}g^{\lambda 0}B_{1}(\omega,{\absvec{k}})+\left(g_{\mu^{\prime}0}g_{\mu 0}\cdot\hat{e}^{\lambda}+\right.\right.\\
&\left.\left.+g_{\mu 0}g^{\lambda 0}\cdot\hat{e}_{\mu^{\prime}}+g_{\mu^{\prime}0}g^{\lambda 0}\cdot\hat{e}_{\mu}\right)B_{2}(\omega,{\absvec{k}})+\left(g_{\mu^{\prime}0}\cdot{\cal P}^{{\rm L}\lambda}_{\mu}(\hat{k})+g_{\mu 0}\cdot{\cal P}^{{\rm L}\lambda}_{\mu^{\prime}}(\hat{k})+g^{\lambda 0}\cdot{\cal P}^{{\rm L}}_{\mu \mu^{\prime}}(\hat{k})\right)B_{3}(\omega,{\absvec{k}})+\right.\\
&\left.+\left[g_{\mu^{\prime} 0}\cdot{\cal P}^{{\rm T}\lambda}_{\mu}(\hat{k})+g_{\mu 0}\cdot{\cal P}^{{\rm T}\lambda}_{\mu^{\prime}}(\hat{k})+g^{\lambda 0}\cdot{\cal P}^{{\rm T}}_{\mu \mu^{\prime}}(\hat{k})\right]B_{4}(\omega,{\absvec{k}})+\hat{e}_{\mu^{\prime}}\hat{e}_{\mu}\hat{e}^{\lambda}B_{5}(\omega,{\absvec{k}})\right.\\
&\left.+\left[\hat{e}_{\mu}\cdot{\cal P}^{{\rm T}\lambda}_{\mu^{\prime}}(\hat{k})+\hat{e}_{\mu^{\prime}}\cdot{\cal P}^{{\rm
T}\lambda}_{\mu}(\hat{k})+\hat{e}^{\lambda}\cdot{\cal P}^{{\rm T}}_{\mu\mu^{\prime}}(\hat{k})\right]B_{6}(\omega,{\absvec{k}})\right]
\end{split}
\end{equation}
\begin{equation}
\begin{split}
&P_{51}=-\dfrac{2 \imath\sqrt{\pi} e^{4} l}{(\pi)^{2}} \langle F_{0}^{\mu \nu}F_{0}^{\mu^{\prime} \nu^{\prime}}\rangle\left(k^{\rho} A^{\sigma}-k^{\sigma}A^{\rho}\right)g_{\sigma 0} \delta_{\nu}^{\lambda}k_{\nu^{\prime}}\left[g_{\mu 0}g_{\mu^{\prime}0}\hat{e}_{\rho}\cdot C_{2}(\omega,{\absvec{k}})+\left(g_{\mu^{\prime}0}\cdot{\cal P}^{{\rm L}}_{\mu \rho}(\hat{k})+\right.\right.\\
& g_{\mu 0}\cdot{\cal P}^{{\rm L}}_{\mu^{\prime} \rho}(\hat{k})C_{3}(\omega,{\absvec{k}})+\left[g_{\mu^{\prime} 0}\cdot{\cal P}^{{\rm T}}_{\mu \rho}(\hat{k})+g_{\mu 0}\cdot{\cal P}^{{\rm T}}_{\mu^{\prime} \rho}(\hat{k})\right]C_{4}(\omega,{\absvec{k}})+\hat{e}_{\rho}\hat{e}_{\mu}\hat{e}_{\mu^{\prime}}\cdot C_{5}(\omega,{\absvec{k}})+\\
&+\left.\left[\hat{e}_{\mu}\cdot{\cal P}^{{\rm T}}_{\mu^{\prime} \rho}(\hat{k})+\hat{e}_{\mu^{\prime}}\cdot{\cal P}^{{\rm T}}_{\mu
\rho}(\hat{k})+\hat{e}_{\rho}\cdot{\cal P}^{{\rm T}}_{\mu \mu^{\prime}}(\hat{k})\right]C_{6}(\omega,{\absvec{k}}) \right]
\end{split}
\end{equation}
\begin{equation}
\begin{split}
&P_{52}=\dfrac{2  e^{4} l^{2}}{(\pi)^{2}} \langle F_{0}^{\mu \nu}F_{0}^{\mu^{\prime} \nu^{\prime}}\rangle\left(k^{\rho} A^{\sigma}-k^{\sigma}A^{\rho}\right)g_{\sigma 0} \delta_{\nu}^{\lambda}k_{\nu^{\prime}}\left[g_{\mu 0}g_{\mu^{\prime}0}\hat{e}_{\rho}\cdot B_{2}(\omega,{\absvec{k}})+\left(g_{\mu^{\prime}0}\cdot{\cal P}^{{\rm L}}_{\mu \rho}(\hat{k})+\right.\right.\\
& g_{\mu 0}\cdot{\cal P}^{{\rm L}}_{\mu^{\prime} \rho}(\hat{k})B_{3}(\omega,{\absvec{k}})+\left[g_{\mu^{\prime} 0}\cdot{\cal P}^{{\rm T}}_{\mu \rho}(\hat{k})+g_{\mu 0}\cdot{\cal P}^{{\rm T}}_{\mu^{\prime} \rho}(\hat{k})\right]B_{4}(\omega,{\absvec{k}})+\hat{e}_{\rho}\hat{e}_{\mu}\hat{e}_{\mu^{\prime}}\cdot B_{5}(\omega,{\absvec{k}})+\\
&+\left.\left[\hat{e}_{\mu}\cdot{\cal P}^{{\rm T}}_{\mu^{\prime} \rho}(\hat{k})+\hat{e}_{\mu^{\prime}}\cdot{\cal P}^{{\rm T}}_{\mu
\rho}(\hat{k})+\hat{e}_{\rho}\cdot{\cal P}^{{\rm T}}_{\mu \mu^{\prime}}(\hat{k})\right]B_{6}(\omega,{\absvec{k}}) \right]
\end{split}
\end{equation}
\begin{equation}
\begin{split}
&P_{61}=-\dfrac{2 \imath\sqrt{\pi} e^{4} l}{(\pi)^{2}} \langle F_{0}^{\mu \nu}F_{0}^{\mu^{\prime} \nu^{\prime}}\rangle\left(k^{\rho} A^{\sigma}-k^{\sigma}A^{\rho}\right)g_{\sigma 0} k_{\nu^{\prime}} k_{\nu^{\prime}}\left[g_{\mu 0}g_{\mu^{\prime} 0}g^{\lambda 0}\hat{e}_\rho D_{2}(\omega,{\absvec{k}})+\left(g_{\mu 0}g^{\lambda 0}\cdot{\cal P}^{{\rm L}}_{\mu^{\prime} \rho}(\hat{k})+\right.\right.\\
&\left.\left.+g_{\mu^{\prime} 0}g^{\lambda 0}\cdot{\cal P}^{{\rm L}}_{\mu \rho}(\hat{k})+g_{\mu 0}g_{\mu^{\prime} 0}\cdot{\cal P}^{{\rm L} \lambda}_{\rho}(\hat{k})\right)D_{3}(\omega,{\absvec{k}})+\left[g_{\mu 0}g^{\lambda 0}\cdot{\cal P}^{{\rm T}}_{\mu^{\prime} \rho}(\hat{k})+g_{\mu^{\prime} 0}g^{\lambda 0}\cdot{\cal P}^{{\rm T}}_{\rho \mu}(\hat{k})+\right.\right.\\
&\left.\left.+g_{\mu 0}g_{\mu^{\prime} 0}\cdot{\cal P}^{{\rm T} \lambda}_{\rho}(\hat{k})\right]D_{4}(\omega,{\absvec{k}})+\left(g_{\mu
0}\cdot\hat{e}_{\rho}\hat{e}_{\mu^{\prime}}\hat{e}^{\lambda}+g_{\mu^{\prime} 0}\cdot\hat{e}_{\rho}\hat{e}_{\mu}\hat{e}^{\lambda}+g^{\lambda
0}\cdot\hat{e}_{\rho}\hat{e}_{\mu^{\prime}}\hat{e}_{\mu}\right)
D_{5}(\omega,{\absvec{k}})+\right.\\
&\left.\left.\left.+g_{\mu 0}\left[\hat{e}^{\lambda}\cdot{\cal P}^{{\rm T}}_{\mu^{\prime} \rho}(\hat{k})+\hat{e}_{\mu^{\prime}}\cdot{\cal P}^{{\rm T} \lambda}_{\rho}(\hat{k})+\hat{e}_{\rho}\cdot{\cal P}^{{\rm T} \lambda}_{\mu^{\prime}}(\hat{k})\right]+g_{\mu^{\prime} 0}\left[\hat{e}^{\lambda}\cdot{\cal P}^{{\rm T}}_{\rho \mu}(\hat{k})+\hat{e}_{\mu}\cdot{\cal P}^{{\rm T} \lambda}_{\rho}(\hat{k})+\hat{e}_{\rho}\cdot{\cal P}^{{\rm T} \lambda}_{\mu}(\hat{k}) \right]+\right.\right.\right.\\
&\left.\left.+g^{\lambda 0}\left[\hat{e}_{\mu^{\prime}}\cdot{\cal P}^{{\rm T}}_{\rho \mu}(\hat{k})+\hat{e}_{\mu}\cdot{\cal P}^{{\rm T}}_{\rho \mu^{\prime}}(\hat{k})+\hat{e}_{\rho}\cdot{\cal P}^{{\rm T}}_{\mu \mu^{\prime}}(\hat{k})\right]\right]D_{6}(\omega,{\absvec{k}})+\hat{e}_{\mu}\hat{e}_{\mu^{\prime}}\hat{e}_{\rho}\hat{e}^{\lambda}D_{7}(\omega,{\absvec{k}})+\left[\hat{e}_{\mu}\hat{e}_{\mu^{\prime}} \cdot{\cal P}^{{\rm T} \lambda}_{\rho}(\hat{k})+\right.\right.\\
&\left.\left.+ \hat{e}_{\mu}\hat{e}_{\rho} \cdot{\cal P}^{{\rm T} \lambda}_{\mu^{\prime}}(\hat{k})+\hat{e}_{\mu}\hat{e}^{\lambda}\cdot{\cal P}^{{\rm T}}_{\mu^{\prime}\rho}(\hat{k})+\hat{e}_{\mu^{\prime}}\hat{e}_{\rho}\cdot{\cal P}^{{\rm T} \lambda}_{\mu}(\hat{k})+\hat{e}_{\mu^{\prime}}\hat{e}^{\lambda}\cdot{\cal P}^{{\rm T} }_{\mu \rho}(\hat{k})+\hat{e}_{\rho}\hat{e}^{\lambda}\cdot{\cal P}^{{\rm T}}_{\mu \mu^{\prime}}(\hat{k})\right]D_{8}(\omega,{\absvec{k}})+\right.\\
&\left.+\left[{\cal P}^{{\rm T}}_{\mu \rho}(\hat{k})\cdot{\cal P}^{{\rm T} \lambda}_{\mu^{\prime}}(\hat{k})+{\cal P}^{{\rm T}
}_{\mu^{\prime}\mu}(\hat{k})\cdot{\cal P}^{{\rm T} \lambda}_{\rho}(\hat{k})+\cdot{\cal P}^{{\rm T} \lambda}_{\mu}(\hat{k})\cdot{\cal P}^{{\rm
T}}_{\mu^{\prime}\rho}(\hat{k})\right]D_{9}(\omega,{\absvec{k}})\right]
\end{split}
\end{equation}
\begin{equation}
\begin{split}
&P_{62}=\dfrac{2 e^{4} l^{2}}{(\pi)^{2}} \langle F_{0}^{\mu \nu}F_{0}^{\mu^{\prime} \nu^{\prime}}\rangle\left(k^{\rho} A^{\sigma}-k^{\sigma}A^{\rho}\right)g_{\sigma 0} k_{\nu^{\prime}} k_{\nu^{\prime}}\left[g_{\mu 0}g_{\mu^{\prime} 0}g^{\lambda 0}\hat{e}_\rho C_{2}(\omega,{\absvec{k}})+\left(g_{\mu 0}g^{\lambda 0}\cdot{\cal P}^{{\rm L}}_{\mu^{\prime} \rho}(\hat{k})+\right.\right.\\
&\left.\left.+g_{\mu^{\prime} 0}g^{\lambda 0}\cdot{\cal P}^{{\rm L}}_{\mu \rho}(\hat{k})+g_{\mu 0}g_{\mu^{\prime} 0}\cdot{\cal P}^{{\rm L} \lambda}_{\rho}(\hat{k})\right)C_{3}(\omega,{\absvec{k}})+\left[g_{\mu 0}g^{\lambda 0}\cdot{\cal P}^{{\rm T}}_{\mu^{\prime} \rho}(\hat{k})+g_{\mu^{\prime} 0}g^{\lambda 0}\cdot{\cal P}^{{\rm T}}_{\rho \mu}(\hat{k})+\right.\right.\\
&\left.\left.+g_{\mu 0}g_{\mu^{\prime} 0}\cdot{\cal P}^{{\rm T} \lambda}_{\rho}(\hat{k})\right]C_{4}(\omega,{\absvec{k}})+\left(g_{\mu
0}\cdot\hat{e}_{\rho}\hat{e}_{\mu^{\prime}}\hat{e}^{\lambda}+g_{\mu^{\prime} 0}\cdot\hat{e}_{\rho}\hat{e}_{\mu}\hat{e}^{\lambda}+g^{\lambda
0}\cdot\hat{e}_{\rho}\hat{e}_{\mu^{\prime}}\hat{e}_{\mu}\right)
C_{5}(\omega,{\absvec{k}})+\right.\\
&\left.\left.\left.+g_{\mu 0}\left[\hat{e}^{\lambda}\cdot{\cal P}^{{\rm T}}_{\mu^{\prime} \rho}(\hat{k})+\hat{e}_{\mu^{\prime}}\cdot{\cal P}^{{\rm T} \lambda}_{\rho}(\hat{k})+\hat{e}_{\rho}\cdot{\cal P}^{{\rm T} \lambda}_{\mu^{\prime}}(\hat{k})\right]+g_{\mu^{\prime} 0}\left[\hat{e}^{\lambda}\cdot{\cal P}^{{\rm T}}_{\rho \mu}(\hat{k})+\hat{e}_{\mu}\cdot{\cal P}^{{\rm T} \lambda}_{\rho}(\hat{k})+\hat{e}_{\rho}\cdot{\cal P}^{{\rm T} \lambda}_{\mu}(\hat{k}) \right]+\right.\right.\right.\\
&\left.\left.+g^{\lambda 0}\left[\hat{e}_{\mu^{\prime}}\cdot{\cal P}^{{\rm T}}_{\rho \mu}(\hat{k})+\hat{e}_{\mu}\cdot{\cal P}^{{\rm T}}_{\rho \mu^{\prime}}(\hat{k})+\hat{e}_{\rho}\cdot{\cal P}^{{\rm T}}_{\mu \mu^{\prime}}(\hat{k})\right]\right]C_{6}(\omega,{\absvec{k}})+\hat{e}_{\mu}\hat{e}_{\mu^{\prime}}\hat{e}_{\rho}\hat{e}^{\lambda}C_{7}(\omega,{\absvec{k}})+\left[\hat{e}_{\mu}\hat{e}_{\mu^{\prime}} \cdot{\cal P}^{{\rm T} \lambda}_{\rho}(\hat{k})+\right.\right.\\
&\left.\left.+ \hat{e}_{\mu}\hat{e}_{\rho} \cdot{\cal P}^{{\rm T} \lambda}_{\mu^{\prime}}(\hat{k})+\hat{e}_{\mu}\hat{e}^{\lambda}\cdot{\cal P}^{{\rm T}}_{\mu^{\prime}\rho}(\hat{k})+\hat{e}_{\mu^{\prime}}\hat{e}_{\rho}\cdot{\cal P}^{{\rm T} \lambda}_{\mu}(\hat{k})+\hat{e}_{\mu^{\prime}}\hat{e}^{\lambda}\cdot{\cal P}^{{\rm T} }_{\mu \rho}(\hat{k})+\hat{e}_{\rho}\hat{e}^{\lambda}\cdot{\cal P}^{{\rm T}}_{\mu \mu^{\prime}}(\hat{k})\right]C_{8}(\omega,{\absvec{k}})+\right.\\
&\left.+\left[{\cal P}^{{\rm T}}_{\mu \rho}(\hat{k})\cdot{\cal P}^{{\rm T} \lambda}_{\mu^{\prime}}(\hat{k})+{\cal P}^{{\rm T}
}_{\mu^{\prime}\mu}(\hat{k})\cdot{\cal P}^{{\rm T} \lambda}_{\rho}(\hat{k})+\cdot{\cal P}^{{\rm T} \lambda}_{\mu}(\hat{k})\cdot{\cal P}^{{\rm
T}}_{\mu^{\prime}\rho}(\hat{k})\right]C_{9}(\omega,{\absvec{k}})\right]
\end{split}
\end{equation}
\subsection{Standard integrals}\label{standint}

In this paragraph we list explicit expressions for the angular integrals $A_{1}\left(\omega,\absvec{k}\right)\ldots
A_{9}\left(\omega,\absvec{k}\right)$, $\ldots$, $D_{1}\left(\omega,\absvec{k}\right)\ldots D_{9}\left(\omega,\absvec{k}\right)$,
$x=\omega/\absvec{k}$:
\begin{equation}
\begin{split}
&A_{1}\left(\omega,\absvec{k}\right)=\int_{0}^{+\pi}d \theta\dfrac{\sin\theta}{(kv)+\imath\epsilon}=\dfrac{1}{|\mathbf{k}|}L[x]\\
&A_{2}\left(\omega,\absvec{k}\right)=\int_{0}^{+\pi}d \theta \dfrac{\cos\theta\sin\theta}{(kv)+\imath\epsilon}=\dfrac{1}{|\mathbf{k}|}(-2+x L[x])\\
&A_{3}\left(\omega,\absvec{k}\right)=\int_{0}^{+\pi}d \theta \dfrac{\cos^{2}\theta\sin\theta}{(kv)+\imath\epsilon}=\dfrac{1}{|\mathbf{k}|}(-2 x+x^{2} L[x])\\
&A_{4}\left(\omega,\absvec{k}\right)=\dfrac{1}{2}(A_{1}\left(\omega,\absvec{k}\right)-A_{3}\left(\omega,\absvec{k}\right))\\
&A_{5}\left(\omega,\absvec{k}\right)=\int_{0}^{+\pi}d \theta \dfrac{\cos^{3}\theta\sin\theta}{(kv)+\imath\epsilon}=\dfrac{1}{|\mathbf{k}|}\left(-\frac{3}{2}-2 x^{2}+x^{3} L[x]\right)\\
&A_{6}\left(\omega,\absvec{k}\right)=\frac{1}{2}(A_{2}\left(\omega,\absvec{k}\right)-A_{5}\left(\omega,\absvec{k}\right))\\
&A_{7}\left(\omega,\absvec{k}\right)=\int_{0}^{+\pi}d \theta \dfrac{\cos^{4}\theta\sin\theta}{(kv)+\imath\epsilon}=\dfrac{1}{|\mathbf{k}|}\left(-\frac{3}{2}x-2 x^{3}+x^{4} L[x]\right)\\
&A_{8}\left(\omega,\absvec{k}\right)=\frac{1}{2}(A_{3}\left(\omega,\absvec{k}\right)-A_{7}\left(\omega,\absvec{k}\right))\\
&A_{9}\left(\omega,\absvec{k}\right)=\frac{1}{4}(A_{1}\left(\omega,\absvec{k}\right)-2A_{3}\left(\omega,\absvec{k}\right)+
A_{7}\left(\omega,\absvec{k}\right))
\end{split}
\end{equation}
\begin{equation}
\begin{split}
&B_{1}\left(\omega,\absvec{k}\right)=\int_{0}^{+\pi}d \theta\dfrac{\sin\theta}{((kv)+\imath\epsilon)^{2}}=-\dfrac{1}{\mathbf{k}^{2}}\dfrac{2}{1-x^{2}}\\
&B_{2}\left(\omega,\absvec{k}\right)=\int_{0}^{+\pi}d \theta\dfrac{\sin\theta\cos\theta}{((kv)+\imath\epsilon)^{2}}=-\dfrac{1}{\mathbf{k}^{2}}\left(L[x]+\dfrac{2x}{1-x^{2}}\right)\\
&B_{3}\left(\omega,\absvec{k}\right)=\int_{0}^{+\pi}d \theta\dfrac{\sin\theta\cos^{2}\theta}{((kv)+\imath\epsilon)^{2}}=-\dfrac{1}{\mathbf{k}^{2}}\left(-2+2x L[x]+\dfrac{2x^{2}}{1-x^{2}} \right)\\
&B_{4}\left(\omega,\absvec{k}\right)=\dfrac{1}{2}(B_{1}\left(\omega,\absvec{k}\right)-B_{3}\left(\omega,\absvec{k}\right))\\
&B_{5}\left(\omega,\absvec{k}\right)=-\dfrac{1}{\mathbf{k}^{2}}\left(-4x+3x^{2}L[x]+\dfrac{2x^{3}}{1-x^{2}} \right)\\
&B_{6}\left(\omega,\absvec{k}\right)=\frac{1}{2}(B_{2}\left(\omega,\absvec{k}\right)-B_{5}\left(\omega,\absvec{k}\right))\\
&B_{7}\left(\omega,\absvec{k}\right)=-\dfrac{1}{\mathbf{k}^{2}}\left(-\dfrac{2}{3}-6x^{2}+4x^{3}L[x]+\dfrac{2x^{4}}{1-x^{2}} \right)\\
&B_{8}\left(\omega,\absvec{k}\right)=\frac{1}{2}(B_{3}\left(\omega,\absvec{k}\right)-B_{7}\left(\omega,\absvec{k}\right))\\
&B_{9}\left(\omega,\absvec{k}\right)=\frac{1}{4}(B_{1}\left(\omega,\absvec{k}\right)-2B_{3}\left(\omega,\absvec{k}\right)+
B_{7}\left(\omega,\absvec{k}\right))
\end{split}
\end{equation}
\begin{equation}
\begin{split}
&C_{1}\left(\omega,\absvec{k}\right)=\int_{0}^{+\pi}d\theta\dfrac{\sin\theta}{((kv)+\imath\epsilon)^{3}}=\dfrac{1}{|\mathbf{k}|^{3}}\dfrac{2x}{(1-x^{2})^{2}}\\
&C_{2}\left(\omega,\absvec{k}\right)=\dfrac{1}{|\mathbf{k}|^{3}}\left(\dfrac{2}{(1-x^{2})}+\dfrac{2x^{2}}{(1-x^{2})^{2}}\right)\\
&C_{3}\left(\omega,\absvec{k}\right)=\dfrac{1}{|\mathbf{k}|^{3}}\left(\dfrac{4x}{(1-x^{2})}+\dfrac{2x^{3}}{(1-x^{2})^{2}}+L[x]\right)\\
&C_{4}\left(\omega,\absvec{k}\right)=\dfrac{1}{2}\left(C_{1}\left(\omega,\absvec{k}\right)-C_{3}\left(\omega,\absvec{k}\right)\right)\\
&C_{5}\left(\omega,\absvec{k}\right)=\dfrac{1}{|\mathbf{k}|^{3}}\left(-2+3x L[x]+\dfrac{6x^{2}}{(1-x^{2})}+\dfrac{2x^{4}}{(1-x^{2})^{2}}\right)\\
&C_{6}\left(\omega,\absvec{k}\right)=\frac{1}{2}(C_{2}\left(\omega,\absvec{k}\right)-C_{5}\left(\omega,\absvec{k}\right))\\
&C_{7}\left(\omega,\absvec{k}\right)=\dfrac{1}{|\mathbf{k}|^{3}}\left(-6 x+6x^{2}L[x]+\dfrac{8 x^{3}}{(1-x^{2})}+\dfrac{2x^{5}}{(1-x^{2})^{2}}\right)\\
&C_{8}\left(\omega,\absvec{k}\right)=\frac{1}{2}(C_{3}\left(\omega,\absvec{k}\right)-C_{7}\left(\omega,\absvec{k}\right))\\
&C_{9}\left(\omega,\absvec{k}\right)=\frac{1}{2}\left(C_{1}\left(\omega,\absvec{k}\right)-2C_{3}\left(\omega,\absvec{k}\right)+
C_{7}\left(\omega,\absvec{k}\right)\right)
\end{split}
\end{equation}
\begin{equation}
\begin{split}
&D_{1}\left(\omega,\absvec{k}\right)=-\dfrac{2}{3\mathbf{k}^{4}}\dfrac{3x^{2}+1}{(1-x^{2})}\\
&D_{2}\left(\omega,\absvec{k}\right)=-\dfrac{1}{|\mathbf{k}|}C_{1}\left(\omega,\absvec{k}\right)+x D_{1}\left(\omega,\absvec{k}\right)\\
&D_{3}\left(\omega,\absvec{k}\right)=-\dfrac{1}{|\mathbf{k}|}(C_{2}\left(\omega,\absvec{k}\right)+x C_{1}\left(\omega,\absvec{k}\right))+
x^{2}D_{1}\left(\omega,\absvec{k}\right)\\
&D_{4}\left(\omega,\absvec{k}\right)=\dfrac{1}{2}(D_{1}\left(\omega,\absvec{k}\right)-D_{3}\left(\omega,\absvec{k}\right))\\
&D_{5}\left(\omega,\absvec{k}\right)=-\dfrac{1}{|\mathbf{k}|}(C_{3}\left(\omega,\absvec{k}\right)+xC_{2}\left(\omega,\absvec{k}\right)+
x^{2}C_{1}\left(\omega,\absvec{k}\right))+x^{3}D_{1}\left(\omega,\absvec{k}\right)\\
&D_{6}\left(\omega,\absvec{k}\right)=\frac{1}{2}(D_{2}\left(\omega,\absvec{k}\right)-D_{5}\left(\omega,\absvec{k}\right))\\
&D_{7}\left(\omega,\absvec{k}\right)=-\dfrac{1}{|\mathbf{k}|}(C_{5}\left(\omega,\absvec{k}\right)+x C_{3}\left(\omega,\absvec{k}\right)+
x^{2}C_{2}\left(\omega,\absvec{k}\right)+x^{3}C_{1}\left(\omega,\absvec{k}\right))+x^{4}D_{1}\left(\omega,\absvec{k}\right)\\
&D_{8}\left(\omega,\absvec{k}\right)=\frac{1}{2}(D_{3}\left(\omega,\absvec{k}\right)-D_{7}\left(\omega,\absvec{k}\right))\\
&D_{9}\left(\omega,\absvec{k}\right)=\frac{1}{2}(D_{1}\left(\omega,\absvec{k}\right)-2D_{3}\left(\omega,\absvec{k}\right)+
D_{7}\left(\omega,\absvec{k}\right))
\end{split}
\end{equation}

\end{document}